\newif\ifnotblinded
\notblindedtrue  % Comment out to hide authors
\documentclass[a4paper, 11pt, onehalfspacing]{article}
\usepackage[fleqn]{amsmath}
\usepackage{mathtools}
\usepackage{amssymb}
\usepackage{setspace}
\usepackage{placeins}
\usepackage{minibox}
\usepackage{graphicx}
\usepackage{tikz}
\usepackage{adjustbox}
\usepackage{color}
\usepackage{colortbl}
\usepackage{xcolor}
\usepackage{caption}
\usepackage{booktabs}
\usepackage{multirow}
\usepackage{authblk}
\usepackage{epstopdf}
\usepackage{algpseudocode}
\usepackage{pgfplots, color, float,bm, lscape,ragged2e}
\usepackage[hmargin = 3cm, vmargin = 2.5cm]{geometry}
\usepackage{natbib}
\usepackage[font=scriptsize]{caption}
\usepackage{float} % to fix table and figure position
\usepackage{booktabs, caption}
\usepackage{subcaption}\usepackage{pgfgantt}
\usepackage{listings}
\usepackage{multicol}
\usepackage{pdfpages}
\usepackage{graphicx} % for picture
\usepackage[toc,page]{appendix}
\usepackage{titlesec}
\titlelabel{\thetitle.\quad}
\usepackage[flushleft]{threeparttable}

\usepackage[labelsep=period]{caption}
\begin{document}
\title{Tail risk forecasting using Bayesian realized EGARCH models}
\ifnotblinded
\author{Vica Tendenan, Richard Gerlach, and Chao Wang}

%\author[1]{Richard Gerlach}
%\author[2]{Vica Tendenan}
%\author[3]{Chao Wang}

\affil{Discipline of Business Analytics, Business School,\\
The University of Sydney, Australia.}
%\affil[1]{The University of Sydney, Australia}
%\affil[2]{The University of Sydney, Australia}

\fi
%\date{.. June 2020}
\date{}
\maketitle

\begin{abstract}

This paper develops a Bayesian framework for the realized exponential generalized autoregressive conditional heteroskedasticity (realized EGARCH) model, which can incorporate multiple realized volatility measures for the modelling of a return series. The realized EGARCH model is extended by adopting a standardized Student-\textit{t} and a standardized skewed Student-\textit{t} distribution for the return equation. Different types of realized measures, such as sub-sampled realized variance, sub-sampled realized range, and realized kernel, are considered in the paper. The Bayesian Markov chain Monte Carlo (MCMC) estimation employs the robust adaptive Metropolis algorithm (RAM) in the burn in period and the standard random walk Metropolis in the sample period. The Bayesian estimators show more favourable results than maximum likelihood estimators in a simulation study. We test the proposed models with several indices to forecast one-step-ahead Value at Risk (VaR) and Expected Shortfall (ES) over a period of 1000 days. Rigorous tail risk forecast evaluations show that the realized EGARCH models employing the standardized skewed Student-\textit{t} distribution and incorporating sub-sampled realized range are favored, compared to a range of models.

\vspace{0.5cm}
\end{abstract}

\begin{onehalfspace}
\section{Introduction}
\label{sec:intro}

Value at Risk (VaR) has become a standard measure in assessing market risks since the introduction of the RiskMetric model by J.P. Morgan in 1993. It is a measure of the maximum potential loss of certain portfolio over a specified holding period at a given confidence level and has been widely employed by corporations and financial institutions in risk management due to its simplicity and convenience in computation. However, the VaR suffers from several weaknesses. First, it provides no information about the size of the losses beyond a particular quantile. Second, it is considered as an incoherent measure of market risks since it violates subadditivity property \citep{McNeilEtAl2005}.

Expected Shortfall (ES) or Conditional VaR was proposed by \citet{ArtznerEtAl1997, ArtznerEtAl1999} as a coherent risk measure to overcome the shortcomings of VaR. It measures the conditional expectation of the return that is higher than a VaR threshold. It can be viewed as an extension of VaR by exploring further details of the tail distribution and therefore can provide a measure about the magnitude of the loss of tailed events, which the VaR cannot provide \citep{YamaiYoshiba2005}. Since May 2012, \citet{Basel2012} has recommended the use of ES to better capture the risks of fat-tailed events. The Basel III Accord, which was implemented in 2019, places new emphasis on ES: ``ES must be computed on a daily basis for the bank-wide internal models to determine market risk capital requirements. ES must also be computed on a daily basis for each trading desk that uses the internal models approach (IMA).'' \citep{BIS2019}.

The calculation of parametric VaR and ES requires an accurate volatility estimation. The parametric approach includes the use of volatility models that estimate conditional quantiles by using the resulted forecasts of conditional volatility. The popular models to capture volatility dynamics are the Autoregressive Conditional Heteroskedasticity (ARCH) model \citep{Engle1982} and the Generalized ARCH model \citep{Bollerslev1986}. The next development of GARCH-class models includes the EGARCH model \citep{Nelson1991} and the GJR-GARCH model \citep{GlostenEtAl1993} that capture the asymmetry in volatility in response to positive and negative returns. These early models utilize daily returns in modelling the conditional variance, which provides noisy signals about the current level of volatility \citep{HansenEtAl2012}. In addition, these models have a slow adjustment to volatility movements, since they rely on long and slowly decaying weighted moving averages of past squared returns \citep{AndersenEtAl2003}. With the availability of high-frequency data, more efficient realized measures of volatility have been proposed, such as realized variance \citep{AndersenBollerslev1998, AndersenEtAl2003}, realized range \citep{MartensVanDijk2007, ChristensenPodolskij2007}, and realized kernel \citep{BarndorffNielsenShephard2002, BarndorffNielsenEtAl2008}.

The realized GARCH model was proposed by \citet{HansenEtAl2012} as an extension of the GARCH-type models, which takes advantage of the availability of the high-frequency data by jointly model the returns and realized measures of volatility. The key feature of the realized GARCH model is an additional equation, called the measurement equation, which contemporaneously relates the conditional variance with a realized measure. Recently, \citet{HansenHuang2016} proposed the realized EGARCH model to improve the realized GARCH model by allowing the use of multiple realized volatility measures and facilitating a more flexible modeling of the dependence between returns and volatility. Another extension of the realized GARCH model, called threshold realized GARCH model, was proposed by \citet{ChenWatanabe2019} by incorporating high-frequency realized volatility measures into a threshold framework to model different volatility behaviors across regimes.

Error distribution specification in the volatility models is also an important aspect of the estimation of the parametric tail risks. Early financial time series modelling based on ARCH and GARCH type models were largely based on the assumption of normal errors, while vast evidence have shown that financial time series are highly leptokurtic and oftentimes skewed. For instance, \citet{Bollerslev1987} allowed for conditional \textit{t}-distributed errors in ARCH and GARCH models; \citet{Hansen1994} and \citet{FernandezSteel1998} proposed a skewed Student-\textit{t} distribution; \citet{ChenEtAl2012} combined an asymmetric Laplace distribution (ALD) with a GJR-GARCH model; \citet{ChenGerlach2013} developed a more flexible extension of the ALD, a two-sided Weibull distribution, as the conditional return distribution in four GARCH-type models.

The original realized GARCH model \citep{HansenEtAl2012} and the realized EGARCH model \citep{HansenHuang2016} both assume Gausiannity of both observation and measurement equation errors. While it is adequate to adopt the normal distribution for measurement equation errors, the study of \citet{Watanabe2012} and \citet{ContinoGerlach2017} on the realized GARCH model suggests that the choice of observation error distribution is found to be important, which in most cases the skewed Student-\textit{t} distribution is favored. \citet{WangEtAl2019} also extended the realized GARCH model by proposing the two-sided Weibull distribution developed by \citet{ChenGerlach2013} for the observation equation errors.

The common method to estimate ARCH and GARCH type models is by using the Maximum Likelihood (ML) estimation. However, the ML estimation faces several challenges. First, since the optimization procedure is usually encumbered by the positivity constraints and stationarity condition, the algorithm can be less stable and produces less consistent results, which will then create problems in standard error calculation \citep{SilvapulleSen2011}. This problem might be more severe for complex models such as the realized EGARCH with multiple realized measures. Second, for the realized EGARCH models, the asymptotic analysis and properties of the ML estimators are complicated and have not been well studied. %Since GARCH-type models are highly non-linear, a very large number of data will be required for the asymptotic argument to hold, which is not always the case in practice  \citep{ArdiaHoogerheide2010}. 
For simplification, \citet{HansenHuang2016} used a score function and the numerical Hessian matrix of the log likelihood function to compute the standard errors for the estimators.

It is argued that the application of the Bayesian approach in the GARCH-type modelling will lead to a better inference and prediction compared to the classical ML approaches. First, the parameter constraints and the the covariance stationarity condition in the GARCH-type models can be incorporated by specifying appropriate priors. Second, the inference of the GARCH-type model parameters is straightforward in the Bayesian setting. Appropriate Markov chain Monte Carlo (MCMC) procedures will allow us to obtain the posterior distribution of any function of the GARCH-type model parameters. The issue of local maxima convergence or convergence to incorrect values which is usually encountered in the ML estimation of complex GARCH-type models can be avoided by using the MCMC techniques. Third, the Bayesian estimation could provide more reliable and consistent results even for small finite samples \citep{ArdiaHoogerheide2010,VirbickaiteEtAl2015}. Fourth, Bayesian approach presents a natural way to assess parameter uncertainty in volatility estimation that can be used in decision making, for instance forecasting purposes. The predictive distributions of the future volatilities and the measure of precision for risk measures can be obtained via predictive interval \citep{Paap2002,AusinGaleano2007}. Relevant to this, the predictive distribution of risk measures such as the VaR is of paramount importance in risk management \citep{Jorion2001}. %Finally, the complexity of GARCH-type model can be handled by using modern Bayesian computational methods such as MCMC procedures \citep{AusinGaleano2007}.

This paper focuses on forecasting the VaR and ES based on the extension of the realized EGARCH model of \citet{HansenHuang2016}. The extension of the realized EGARCH is conducted in several ways. First, two types of return distributions, i.e. a standardized Student-\textit{t} and a standardized skewed Student-\textit{t} distribution, are proposed, to take into account the non-Gaussianity in the error distribution. The measurement equation errors are kept to be Normal distribution. The two frameworks are called RE-SkN and RE-tN. Second, three types of realized measures are utilized in the realized EGARCH model, either individually or jointly. Those are the sub-sampled realized variance (RVSS), sub-sampled realized range (RRSS), and realized kernel (RK). The sub-sampling process is based on \citet{ZhangEtAl2005} to deal with micro-structure effects. The choice of the RVSS and RRSS is motivated by the recent works of \citet{GerlachWang2016} and \citet{WangEtAl2019} who found that the RVSS and RRSS could improve the out-of-sample forecasting of their proposed models. Meanwhile, the RK is chosen since it has some robustness to market microstructure effects \citep{BarndorffNielsenEtAl2008} and has been consistently used in the original realized GARCH and realized EGARCH models. Third, an adaptive MCMC procedures is adopted and applied to the proposed models. To assess the performance of the proposed RE-SkN and RE-tN models, their 2.5\% and 1\% (following the recommendation of Basel Committee \citeyearpar{Basel2012, Basel2016, BIS2019}) VaR and ES forecasts are compared to those of various competing models such as the original realized GARCH and realized EGARCH models, the Conditional Autoregressive Expectile (CARE) model by \citet{Taylor2008}, and several extensions of GARCH-type models. 

The structure of the paper is as follows: Sections 2 reviews the employed realized measures, Section 3 discusses the proposed extension to the realized EGARCH models, and Section 4 discusses the Bayesian approach to estimate the proposed models. In Section 5, we present the result of a simulation study using the proposed method and models. Section 6 provides an overview of tail risks forecasting methods and relevant model evaluations. Section 7 presents empirical study results, followed by the conclusion of the paper and possible extensions in Section 8.

\section{Realized Measures}
\label{sec:realized_measures}

This section reviews three types of realized measures employed in this study, i.e. the realized variance (RV), realized range (RR), and realized kernel (RK). For day $t$, let $H_{t}$, $L_{t}$ and $C_{t}$ denote the intra-day high, low, and closing prices, respectively. The daily close-to-close log return is calculated as:
\begin{eqnarray}\label{return_def}
r_t= \log(C_t)-\log(C_{t-1}), \nonumber
\end{eqnarray}
\noindent
where $r_t^2$ is the associated volatility estimator.

In a high frequency intra-day framework, each day $t$ from open to close can be divided into $N$ intervals of equal size with length $\triangle$, where each intra-day time is subscripted as $i= 0, 1, 2,...,N$. Let $P_{t-1+i\triangle}$ denotes the closing price at the $i$-$th$ interval of day $t$. RV is calculated as the sum of the $N$ intra-day squared returns, at frequency $\triangle$, for day $t$:
\begin{eqnarray}\label{rv_def1}
RV_{t}^{\triangle}=\sum_{i=1}^{N} \big[\log(P_{t-1+i\triangle})-\log(P_{t-1+(i-1)\triangle})\big]^{2}.
\end{eqnarray}

The RR, proposed by \citet{MartensVanDijk2007} and \citet{ChristensenPodolskij2007}, may contain more information about volatility than the RV, in the same way that the intra-day range contains more information than squared returns, i.e. it uses all the price movements in a time period, not just the price at the start and end of each sub-period. Let $H_{t,i}=\sup_{(i-1)\triangle<j<i\triangle}P_{t-1+j}$, and $L_{t,i}=\inf_{(i-1)\triangle<j<i\triangle}{P_{t-1+j}}$ denote the high and low prices during $i$-$th$ interval of day $t$, respectively, the RR is simply the sum of the the squared intra-period ranges:
\begin{eqnarray}\label{rrv_def}
RR_{t}^{\triangle}= \frac {\sum_{i=1}^{N}(\log H_{t,i}-\log L_{t,i})^2}{4\log2}.
\end{eqnarray}

In an ideal situation where prices are observed continuously and without measurement error, both the RV and RR will generate a perfect estimate of volatility. Unfortunately, the RV suffers from a well-known bias problem due to market microstructure noises. The bias tends to get worse as the sampling frequency of intra-day returns increases \citep{HansenLunde2006} and is evident in volatility signature plots \citep{Fang1996, AndersenEtAl2000}. 

Nevertheless, \citet{MartensVanDijk2007} proved that the RR is more competitive and sometimes more efficient the RV, under some micro-structure conditions and levels. As $N \rightarrow \infty $, the scaling factor of $4\log2$ makes the RR an approximately unbiased estimator. Even though $N$ is finite for empirical data and equation (\ref{rrv_def}) is biased, when employed in the realized GARCH and realized EGARCH models, the measurement equation parameters can adjust for such bias. Therefore, the RR is not required to be unbiased in the realized GARCH and the realized EGARCH models. \citet{GerlachWang2016} proved that that employing the RR as the measurement equation variable in a realized GARCH model can improve predictive likelihoods, accuracy, and efficiency in forecasting empirical tail risks.

There are different approaches that can be adopted to to reduce the market microstructure noise in the RV and RR, such as scaling and subsampling approaches. The scaling process was proposed by \citet{MartensVanDijk2007}  as follows:
\begin{eqnarray}\label{rv_scale}
RV_{S,t}^{\triangle}= \frac {\sum_{l=1}^{q}RV_{t-l}}{\sum_{l=1}^{q}RV_{t-l}^{\triangle}}RV_{t}^{\triangle},
\end{eqnarray}
\begin{eqnarray}\label{rrv_scale}
RR_{S,t}^{\triangle}= \frac {\sum_{l=1}^{q}RR_{t-l}}{\sum_{l=1}^{q}RR_{t-l}^{\triangle}}RR_{t}^{\triangle},
\end{eqnarray}
\noindent
where $RV_{t}$ and $RR_{t}$ represent the daily return square and range square at day $t$, repectively, and $q$ is the number of days employed to estimate the scaling factors. 

The motivation behind the scaling process is the fact that micro-structure noise has less effect on the daily return and range than on their high frequency counterparts. Therefore, the scaling factor can help reducing the bias in the RV and RR via the scaling and smoothing processes.

Meanwhile the sub-sampling approach was proposed by \citet{ZhangEtAl2005}, by which a number of sub grids is selected from the original grid of observation times and then the average of estimators derived from the sub grids is calculated. For day $t$, $N$ equally sized samples are grouped into $M$ non-overlapping subsets $\Theta^{(m)}$ with size $N/M=n_{k}$, which means:
\begin{eqnarray}
\Theta = \bigcup_{m=1}^{M} X^{(m)}, \; \mbox{where} \; \Theta^{(k)}  \cap \Theta^{(l)} = \emptyset, \;  \mbox{when}  \;  k \neq l.  \nonumber
\end{eqnarray}

The sub-sampling process is then implemented on the subsets $\Theta^{(i)}$ with $n_{k}$ interval:
\begin{eqnarray}
\Theta^{(i)}= {i, i+n_k,...,i+n_k(M-2), i+n_k(M-1)}, \; \mbox{where} \;  i= {0,1,2...,n_k-1}.  \nonumber
\end{eqnarray}

\citet{WangEtAl2019} applied the sub-sampling procedures on the RV and RR. Let $C_{t,i}=P_{t-1+i\triangle}$ denotes the log closing price at the $i$-$th$ interval of day $t$, the RV with the subsets $\Theta^{i}$ is calculated as:
\begin{eqnarray}
RV_{i}= \sum_{m=1}^{M} (C_{t,i+n_{k}m}-C_{t,i+n_{k}(m-1)})^{2}; \; \mbox{where} \; i= {0,1,2...,n_k-1}.  \nonumber
\end{eqnarray}

Suppose there are $T$ minutes per trading day, the $T/M$ RV with $T/N$ sub-sampling is given by:
\begin{eqnarray}
RV_{T/M,T/N}= \frac{\sum_{i=0}^{n_k-1} RV_{i} } {n_k}.
\end{eqnarray}

Let $H_{t,i}=\sup_{(i+n_{k}(m-1))\triangle<j<(i+n_{k}m) \triangle}P_{t-1+j}$ and $L_{t,i}=\inf_{(i+n_{k}(m-1))\triangle<j<(i+n_{k}m) \triangle}{P_{t-1+j}}$ denote the high and low prices during the interval $i+n_{k}(m-1)$ and $i+n_{k}m$, respectively. \citet{WangEtAl2019}  proposed the $T/M$ RR with $T/N$ sub-sampling that takes the following form:
\begin{eqnarray}
RR_{i}= \sum_{m=1}^{M} (H_{t,i}-L_{t,i})^2; \, \text{  where } \; i= {0,1,2...,n_k-1}.
\end{eqnarray}
\begin{eqnarray}
RR_{T/M,T/N}= \frac{\sum_{i=0}^{n_k-1} RR_{i} } {4\log2 n_k}.
\end{eqnarray}
\noindent
For details regarding the properties of the sub-sampled RR, please refer to \citet{WangEtAl2019}.

The selection of sampling frequency needs to take into account a trade-off between bias and variance, e.g. see \citet{BandiRussell2008} and \citet{ZhangEtAl2005}. The RV and RR are usually computed from intra-day returns sampled at a moderate frequency, such as 5-minute frequency, to balance the bias and variance. This study employs 5-minute RV and RR with 1 min sub-sampling proposed by \citet{WangEtAl2019}, as follows:
\begin{gather}  \nonumber
RV_{5,1,0}=(\log C_{t5}-\log C_{t0})^2+(\log C_{t10}-\log C_{t5})^2+...\\   \nonumber
RV_{5,1,1}=(\log C_{t6}-\log C_{t1})^2+(\log C_{t11}-\log C_{t6})^2+...\\  \nonumber
RV_{5,1}=\frac{\sum_{i=0}^{4}RV_{5,1,i}} {5},\\  \nonumber
RR_{5,1,0}=(\log H_{t0\leq t \leq t5}-\log L_{t0\leq t \leq  t5})^2+(\log H_{t5\leq t \leq t10}-\log L_{t5\leq t \leq  t10})^2+...\\  \nonumber
RR_{5,1,1}=(\log H_{t1\leq t \leq t6}-\log L_{t1\leq t \leq  t6})^2+(\log H_{t6\leq t \leq t11}-\log L_{t6\leq t \leq  t11})^2+...\\  \nonumber
RR_{5,1}=\frac{\sum_{i=0}^{4}RR_{5,1,i}} {4 \log (2)5}. \nonumber
\end{gather}

The third realized measure proposed to be employed in this study is the realized kernel (RK), which is somewhat robust to noise. The RK was developed by \citet{BarndorffNielsenEtAl2008} and was also employed the original realized GARCH and realized EGARCH models. The estimator takes the following form:
\begin{eqnarray} \label{rk_parzen}
    K(X)=\sum_{h=-H}^{H} k \Big( \frac{H}{H+1} \Big) \gamma_{h}, \; \mbox{where} \;\;\; \gamma_{h}=\sum_{j=\mid h \mid+1}^{n} x_{j,t}x_{j-\mid h \mid,t}.
\end{eqnarray}

The kernel weight function, $k(x)$, in equation (\ref{rk_parzen}), is chosen to be the Parzen weight function, since it satisfies the smoothness conditions, $k'(0) = k'(1) =0 $. This will ensure a non-negative estimate. The parzen kernel function is defined as follows:
\begin{eqnarray} \label{parzen_kernel_fn}
k(x)=
    \begin{cases}
        1-6x^2+6x^3 & 0\geq x \geq 1/2, \\
        2(1-x)^3 & 1/2 \geq x \geq 1, \\
        0 & x > 1,
    \end{cases}
\end{eqnarray}
\noindent
where $x_j$ is the high-frequency return in equation (\ref{rk_parzen}) and (\ref{parzen_kernel_fn}), $H$ is a standard bandwidth spelt out in \citet{BarndorffNielsenEtAl2009}, given by:
\begin{align*}
    & H^*=c^*\xi^{4/5}n^{3/5}, \hspace{0.5cm} c^*=((12)^2/0.2969)^{1/5}=3.5134, \hspace{0.5cm} \xi^2 = \frac{\omega^2}{T \int_0^T \sigma^4_u du}.
\end{align*}
\noindent
For details, please see \citet{BarndorffNielsenEtAl2009}.

\section{Proposed model}
\label{sec:model}
The realized EGARCH model of \citet{HansenHuang2016} with \textit{K} realized measures takes the following form: 
\begin{eqnarray} \label{RealEGARCH}
    r_t &=& \mu + \sqrt{h_t}\epsilon_t, \nonumber\\ 
    \log h_t &=&  \omega + \beta \log h_{t-1} + \tau (\epsilon_{t-1}) + \gamma^{'}u_{t-1}, \nonumber \\
    \log x_{k,t} &=&  \xi_{k} + \varphi_{k}\log h_t + \delta_{(k)}(\epsilon_t) + u_{k,t}, \>\>\> k=1, ..., K,
\end{eqnarray}
where $r_t$ is the daily return, $\epsilon_t \overset{iid}{\sim} D_1(0,1)$ and $ u_t \overset{iid}{\sim}  D_2(0,\Sigma)$, $\epsilon_t $ and $ u_t=(u_{1,t},...,u_{K,t})^{'}$ are mutually and serially independent, $x_t$ is the vector of realized measures, and $\tau(\epsilon)$ and $\delta_{(k)}(\epsilon)$ are leverage functions given by $\tau(\epsilon) = \tau_{1}\epsilon+\tau_{2}(\epsilon^{2}-1)$ and $\delta_{(k)}(\epsilon) = \delta_{k,1}\epsilon+\delta_{k,2}(\epsilon^{2}-1), \>\>\> k=1...,K.$, respectively.

The realized EGARCH model in equation (\ref{RealEGARCH}) consists of three types of equations. The first equation is the standard return equation. The second equation is the GARCH equation and the third type is the measurement equation(s). There are two main features of the GARCH equation in the realized EGARCH model. The first feature is the presence of a leverage function, $\tau(z_{t-1})$ that distinguishes the realized EGARCH model from the realized GARCH model, which only includes a leverage function in the measurement equation. The second key feature is the last term, $\gamma^{'}u_{t-1}$, which facilitates the channelling of the impact of realized measure on the expectations of future volatility. Since $u_t$ is K-dimensional, it allows us to incorporate multiple realized measures of volatility. The parameter $\beta$ represents volatility persistence, whereas $\gamma$ reflects the information the realized measures contain about future volatility. The measurement equation(s) show the relationship between the (\textit{ex-post}) realized measures of volatility and the (\textit{ex-ante}) conditional variance. In their paper, \citet{HansenHuang2016} used the RK, the daily range, and six types of realized variances and adopt a Gaussian specification, $D_1(0,1) = N(0,1)$ and $ D_2(0,\Sigma) =  N(0,\Sigma)$. 

This study extends the realized EGARCH model by allowing $D_1(0,1)$ to be a standardized Student-\textit{t}, and a standardized skewed Student-\textit{t} of \citet{Hansen1994}, following \citet{Watanabe2012} and \citet{ContinoGerlach2017} for the case of the realized GARCH. These proposed models are denoted hereafter as RE-tN and RE-SkN, respectively, while the original realized EGARCH model employing normal distribution for both return and measurement equations is denoted as RE-NN.

The Student-\textit{t} distribution could be potentially restrictive, since it only allows a variation in the location, scale, and tail fatness. Meanwhile, the skewed Student-\textit{t} distribution is more flexible and allows for a richer behaviors. It is able to model not only leptokurtosis, but also asymmetry \citep{Hansen1994}. The pdf of a standardized skewed Student-\textit{t} distribution is: 
\begin{eqnarray} \label{Skt_pdf}
    (\varepsilon|\nu,\lambda)=
    \begin{cases}
        bc\left( 1 + \frac{1}{\nu-2} \left( \frac{b\varepsilon+a}{1-\lambda} \right)^2 \right)^{-(\nu+1)/2} & if \hspace{0.5cm}  \varepsilon < -\frac{a}{b},\\
        bc\left( 1 + \frac{1}{\nu-2} \left( \frac{b\varepsilon+a}{1+\lambda} \right)^2 \right)^{-(\nu+1)/2} & if \hspace{0.5cm} \varepsilon \geq -\frac{a}{b},\\
    \end{cases}
\end{eqnarray}
\noindent
where $\nu$ is the degrees of freedom parameter that controls the kurtosis of the distribution with range $2 < \nu < \infty$, $\lambda$ is the skewness parameter with range $-1 <\lambda < 1$, while $a,b,$ and $c$ are some constants given by:
\begin{eqnarray}
a=4 \lambda c \left( \frac{\nu-2}{\nu-1} \right), \hspace{0.5cm} b=\sqrt{1+3\lambda^2-a^2}, \hspace{0.5cm} c=\frac{\Gamma \Big( \frac{\nu+1}{2} \Big)}{\sqrt{\pi(\nu-2)}\Gamma \Big( \frac{\nu}{2} \Big)}. \nonumber
\end{eqnarray}

\citet{Hansen1994} showed that this is a proper density function that has zero mean and a unit variance and specializes to a Student-\textit{t} density by setting $\lambda = 0$. When $\lambda > 0$, the mode of the density is to the left of zero and the variable is right-skewed, and vice versa when $\lambda < 0$.

The stationarity condition of the realized EGARCH model in equation (\ref{RealEGARCH}) can be derived by subtituting the measurement equation into the GARCH equation:
\begin{eqnarray} \label{GARCH_eq}
    \log h_t = \omega - \gamma^{'}\xi_k +(\beta-\gamma^{'}\varphi_k)\log h_{t-1} + a_t,
\end{eqnarray}
\noindent
where $a_t = \tau(\epsilon_{t-1}) + \gamma^{'}(\log x_{k,t-1} - \delta_{(k)}(\epsilon_t))$. 

Taking the expectation of both sides of equation (\ref{GARCH_eq}), we derive the long run value of $\log h$ as given by:
\begin{eqnarray}
    E(\log h)=\frac{\omega - \gamma^{'}\xi_k}{1-(\beta-\gamma^{'}\varphi_k)}. 
\end{eqnarray}

The stationarity of the realized EGARCH model in equation (\ref{RealEGARCH}) is then achieved by imposing the following restriction:
\begin{eqnarray} \label{stationarity}
\beta-\gamma^{'}\varphi_k < 1.
\end{eqnarray}

\section{Bayesian Estimation}
This section discusses the Bayesian estimation approach and Markov chain Monte Carlo (MCMC) sampling procedures employed in estimating the proposed models and conducting the tail-risk forecasting. Basically, the Bayesian model is a probability model that consists of a likelihood function $p(y|\theta) $ and a prior distribution $\pi(\theta)$. The product of the likelihood and the prior probability is equivalent to the posterior distribution $p(\theta|y)$. 

\subsection{Likelihood}

The log-likelihood function for the RE-SkN type models, where $D_1 \sim Skt_{\nu,\lambda}(0, 1)$ and $D_2 = N(0,\Sigma)$ is given by: 

\begin{eqnarray} \label{RealEGARCH_llfunction-Skt}
    \ell(r; \theta)=
    \begin{cases}
     T [\ell(b) + \ell(c)] - \sum_{t=1}^{T} \Bigg[ \frac{v+1}{2} \log \Bigg(1+\frac{1}{v-2}\Bigg(\frac{b \epsilon_{t}+a}{1-\lambda}\Bigg)^2\Bigg)+0.5\log(h_t)] & \mbox {if $\epsilon_{t} < -\frac{a}{b}$}, \\
    \cr
    T [\ell(b) + \ell(c)] - \sum_{t=1}^{T} \Bigg[ \frac{v+1}{2} \log \Bigg(1+\frac{1}{v-2}\Bigg(\frac{b \epsilon_{t}+a}{1+\lambda}\Bigg)^2\Bigg)+0.5\log(h_t)] & \mbox {if $\epsilon_{t} \geq -\frac{a}{b}$}. \nonumber
    \end{cases}
\end{eqnarray}

\begin{eqnarray}
\ell(x;\theta,\Sigma) = -\frac{1}{2}\sum\limits_{t=1}^{n} K\log(2\pi) + \log(|\Sigma|) + u^{'}_{t}\Sigma^{-1}u_t], \nonumber
 \end{eqnarray}
\begin{eqnarray}  
 \ell(r,x;\theta,\Sigma) = \ell(r; \theta) + \ell(x;\theta,\Sigma).
 \end{eqnarray}

Meanwhile, the log-likelihood function for the RE-tN type model, where $D_1 \sim t_{\nu}(0, 1)$ and $D_2 = N(0,\Sigma)$ takes the following form: 
\begin{eqnarray} \label{RealEGARCH_llfunction-tN}
    \begin{split}
    \ell(r,x;\theta,\Sigma) &= T \Big[\log\Gamma \Big(\frac{v+1}{2} \Big) - \log \Gamma \Big(\frac{v}{2}\Big) - 0.5 \log[(v-2) \pi]\Big] \cr
    &- \sum_{t=1}^{T} \Bigg[\frac {v+1}{2} \log \Bigg(1 + \frac{(r_{t}-\mu)^2}{(v-2)h_t}\Big) + 0.5\log h_t \Big] \cr
    & -\frac{1}{2}\sum\limits_{t=1}^{n} \Big[ K\log(2\pi) + \log(|\Sigma|) + u^{'}_{t}\Sigma^{-1}u_t \Big].
    \end{split}
\end{eqnarray}

The parameters in the realized EGARCH models are:
\begin{eqnarray} \label{RealEGARCH_params}
    \theta = (\mu ,\omega ,\beta, \tau^{'},\gamma^{'}, \psi^{'}_{1},...,\psi^{'}_{K},\nu, \lambda)^{\prime} \>\> \mathrm{and} \>\> \Sigma,
\end{eqnarray}
    where $\psi_k=(\xi_k,\varphi_k, \delta^{'}_{k})^{\prime}, \>\> k=1,...,K.$\\

Joint Conditional Likelihood is given by:
    \begin{eqnarray} \label{Joint_Cond_Likelihood}
    \mathcal{L} (\theta,\Sigma,r,x) = \displaystyle\prod_{t=1}^{T} p(r_t,x_t|\mathcal{F}_{t-1},\theta, \Sigma) \\
    = \displaystyle\prod_{t=1}^{T} p(r_t|\mathcal{F}_{t-1},\theta) p(x_t|r_t,\mathcal{F}_{t-1},\theta, \Sigma)
    \end{eqnarray}
    where $\mathcal{F}_t=(r_1,...,r_t;x_1,...,x_t)'$; the predictive density $p(r_t \mid \mathcal{F}_{t-1},\theta)$ is determined by the distributional choice of $\epsilon_t$; and the conditional density $p(x_t|r_t,\mathcal{F}_{t-1},\theta, \Sigma)$ is determined by the distribution of $u_t$.

\subsection{Prior Specification}
A combination of mostly uninformative and Jeffreys-type priors is chosen over the possible region for the parameters:
\begin{eqnarray} \label{prior}
\pi(\theta) \propto \frac{1}{\sigma_{u,k=1}^2} \frac{1}{\sigma_{u,k=2}^2} .. \frac{1}{\sigma_{u,k=K}^2} \frac{1}{\nu^2}I(A),
\end{eqnarray}
where A is the region described by the parameter restriction in Table \ref{table:parameter_restriction}. The indicator function $I(A)=1$ over the region A and zero otherwise.

\begin{table}[H]
    \captionsetup{font=normalsize, skip=0pt, justification= centering}
    \caption {Parameter Restriction} \label{table:ParameterBound} 
    \label{table:parameter_restriction}
    \begin{center}
    \begin{tabular}{|l|c|}
    \hline
    Parameter                                               & Bounds        \\ \hline
    $\mu, \omega,\tau_1, \tau_2, \xi, \delta_1, \delta_2 $  & $\pm \infty$  \\ \hline
    $\sigma^2_u$                                            & $ > 0 $       \\ \hline
    $\nu$                                               & $ (4,200) $       \\ \hline
    $\lambda$                                            & $ (-1,1) $       \\ \hline
    $\beta-\gamma^{'}\varphi$                            & $ < 1 $     \\ \hline
    \end{tabular}
    \end{center}
\end{table}

We put a standard Jeffreys prior on the variance of $u_t$ in Equation (\ref{RealEGARCH}). 
The prior of the degrees of freedom parameter ($\nu$) is $\nu^{-2}$, which is an improper prior obtained by being flat on $\nu^{-1}$ \citep{BauwensEtAl2006}. This is equivalent to $U(0,\frac{1}{4})$ so that $\nu > 4$. This ensures that the first 4 moments of the distribution of $\epsilon_t$ are finite \citep{ChenEtAl2006}. Other research works have adopted this type of prior on $\sigma^2_u$ and $\nu$ on Bayesian estimation of the previous model, the realized GARCH models, e.g. see \citet{GerlachWang2016}, \citet{ContinoGerlach2017}, and \citet{WangEtAl2019}.

\subsection{Adaptive MCMC}
\label{sec:mcmc}
In order to numerically perform an inference on the model parameters, a MCMC procedure is performed to draw samples from the posterior distribution of the model parameters. An adaptive MCMC method of \citet{ContinoGerlach2017}
is adopted and extended, by considering the balancing between the target acceptance rate and convergence, the scaling factor for the tuning process, and the selection of the block of the parameters in the updating process. The adaptive MCMC approach consists of two steps, burn-in period and post burn-in period.

In the burn-in period, a Robust Adaptive Metropolis (RAM) algorithm of \citet{Vihola2012} is employed. The RAM algorithm estimates the shape of the target distribution and simultaneously allow us to attain a given mean acceptance rate. The RAM is defined recursively through the following process:
\begin{enumerate}
    \item Compute the proposal $Y_n := X_{n-1}+S_{n-1}U_n$, where $S_{n-1}$ is a lower diagonal matrix with positive diagonal elements and $U_n$ is a vector of random variable from $d$-dimensional Gaussian distribution.
    \item The proposal is accepted with probability $\alpha_n:=\min\left\{1,\frac{\pi(Y_n)}{\pi(X_{n-1})} \right\}$.
    \item If the proposal $Y_n$ is accepted, set $X_n:= Y_n$. Otherwise, set $X_n:=X_{n-1}$.
    \item Compute the lower diagonal matrix, $S_n$, satisfying the equation: 
        \begin{eqnarray} \label{RAM_lower_diagmat}
        S_nS_n^T=S_{n-1} \Big( I+\eta_n (\alpha_n-\alpha^*)\frac{U_nU_n^T}{||U_n||^2} S_{n-1}^T \Big),         \end{eqnarray}
        where $ \big \{\eta_n \big \}_{n \geq 1} \subset (0,1]$ is a step size sequence decaying to zero, often defined as $\eta_n = n^{-\gamma}$ with an exponent $\gamma \in (1/2,1)$, $\alpha^* \in (0,1)$ is the target mean acceptance probability, and $I \in \mathbb{R}^{dxd}$ is the identity matrix. 
    \end{enumerate}
    
To obtain $S_n$ in step (4), we use rank-one Cholesky update of $S_{n-1}$ when $(\alpha_n-\alpha^*) > 0$ or downdate of $S_{n-1}$ when $(\alpha_n-\alpha^*) < 0$. This approach is computationally more efficient than computing the right hand side of equation (\ref{RAM_lower_diagmat}) and performing new Cholesky decomposition. Folowing \citet{Vihola2012}, $\eta_n = \min \{1,d \cdot n^{-2/3}\}$, where $d$ is the dimension of the block of parameters. The target mean acceptance rate $\alpha^* = 0.234$ (if $d>4$), or 0.35  (if $2 \leq d \leq 4)$, or 0.44 (if $d = 1$), as recommended by \citet{RobertsEtAl1997}. For theoretical details of the RAM algorithm, see \citet{Vihola2012}.

The algorithm is run for 20,000 iterations for each model parameter to achieve convergence. The first half of burn-in iterates (10,000 iterates) is discarded, then the remaining second half is used to calculate the sample mean ($X_n$) and covariance matrix ($\Sigma$) for each parameter as the proposal distribution in the post burn-in period. In the post burn-in period, a standard RWM algorithm is employed and a mixture of three Gaussian proposal distribution is used as follows: 
\begin{eqnarray}
 Y_n \sim \mathcal{N} \big(X_{n-1},C_i\Sigma \big),
\end{eqnarray}
where $C_1 = 1; C_2 = 100, C_3 = 0.01$, following \citet{WangEtAl2019}.
The algorithm is run for 10,000 iterations for each model parameter, which are then used to calculate the tail risk forecasts. The posterior mean of the tail risk forecasts is used as the final forecast.

The convergence of the adaptive MCMC method is assessed via the comparison of the estimated variances between-chains and within-chains for each model parameter by using the Gelman-Rubin diagnostic. The convergence is indicated by small differences between these variances \citep{GelmanEtAl2013}. The efficiency of the MCMC method is evaluated by using an effective sample size testing \citep{GelmanEtAl2013} and the autocorrelation
time \citep{ChibJeliazkov2001}.

To determine the setting of the parameter blocks, two settings of parameter blocks are compared: 1 block (as in \citet{ContinoGerlach2017})  and 4 blocks. The latter setting  
is adapted from \cite{WangEtAl2019} and adjusted to the proposed models in this paper as follows: $\theta_1 = (\mu$), $\theta_2 = (\omega, \beta,\tau', \gamma', \varphi_k$),  $\theta_3 = (\xi_k,\delta'_k, \Sigma$), and $\theta_4 = (\nu, \lambda$). The parameters within the same parameter block are more likely to be correlated, in the posterior (or likelihood), than those between blocks. For example, the stationarity condition causes correlation between iterates of $\beta,\gamma', \varphi_k$. Therefore, those parameters are are grouped in the same block.

\section{Simulation study}
\label{sec:simulation}
A simulation study is conducted to compare the performance of the MCMC estimators and maximum likelihood (ML) estimators of the proposed RE-SkN model. The purpose is to demonstrate the superior performance of the MCMC estimators in terms of unbiasedness and precision properties. 

A total of 1000 replicated datasets are simulated from the proposed RE-SkN model using 1 realized measure, with a sample size of 2000. 
\begin{eqnarray} \label{RealEGARCH_sim1}
r_t & = & 0.0+ \sqrt{h_t}\epsilon_t \\ \notag
\log h_t & = & -0.12 + 0.98\log h_{t-1} -0.12\epsilon_{t-1} + 0.04(\epsilon^2_{t-1}-1) + 0.47u_{t-1} \\ \notag
\log x_{t} & = & -0.17 + 0.94\log h_t -0.09\epsilon_t + 0.06(\epsilon_{t}^2-1) + u_{t}, 
\end{eqnarray}
where  $\epsilon_{t} \sim Skt_{\nu=4.4, \lambda=0.5}$, $u_{t} \sim N(0, 0.15)$, and $h_0 = 0.0025$. 

All initial parameter values were arbitrarily set equal to 0.1, except for $\nu=5$, for both estimation methods. The tail risks are considered for the quantile level $\alpha=2.5\%$ and $\alpha=1\%$, following the recommendation of Basel Committee \citeyearpar{Basel2012, Basel2016,BIS2019}. 

\begin{figure}[H]
    \begin{center}
        \captionsetup{font=normalsize, skip=0pt, justification= centering}
        \caption{MCMC convergence and effficiency using 4 parameter blocks setting}
        \label{fig:MCMC_convergence}
        \includegraphics[width=0.9\linewidth]{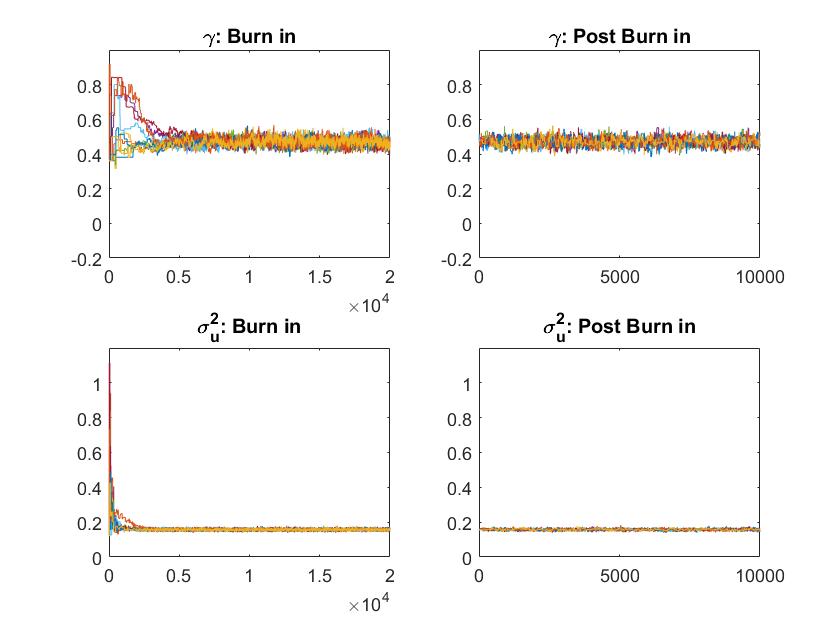}
    \end{center}
\end{figure}

First, to assess the convergence of the MCMC method, $m=10$ independent MCMC chains are run for two different parameter block settings. Then, we analyze the variance between and within the chains according to the Gelman-Rubin diagnostic \citep{GelmanEtAl2013}. Table \ref{table:convergence} presents the summary of the the Gelman-Rubin diagnostic $\hat{R}$, the number of effective sample size $n_{\text{eff}}$ \citep{GelmanEtAl2013}, and the autocorrelation time $\hat{t}$ \citep{ChibJeliazkov2001} for both 1 parameter block and 4 parameter blocks settings. Favorable results are indicated by the values printed in bold. The $\hat{R}$ for each parameter is very close to 1 and all $\hat{R}$ are $<1.1$, the recommended threshold by \citet{GelmanEtAl2013}, except for $\nu$ under the 1 block setting. The $\hat{R}$ under the 4 blocks settings is closer to 1 for 11 out of 13 parameters, suggesting an improved convergence property under this parameter block setting. The convergence can also be assessed visually by using the trace plots of the MCMC iterations of those independent chains, as depicted for $\gamma$ and $\sigma_u^2$ parameters in Figure \ref{fig:MCMC_convergence}. As can be seen, the MCMC chains show a rapid convergence and chain mixing.

The result also suggests that the number of effective sample size for most parameters, $n_{\text{eff}}$, is higher than the suggested benchmark ($5m$) by \citet{GelmanEtAl2013}. The number of effective sample size for 10 parameters under the 4 blocks setting is higher than that under the 1 block setting. As regards the correlation time $\hat{t}$, the parameters under the 4 blocks setting also have a more favorable property, i.e. have a smaller autocorrelation time. Overall, the MCMC method under the 4 parameter blocks setting provides a much more improved convergence property, is more efficient, and has a smaller autocorrelation time than the 1 block setting. 

\begin{table}[H]
\begin{center}
  %\captionsetup{font=scriptsize}
  \captionsetup{font=normalsize, skip=0pt, justification= centering}
  \caption{Summary statistics of the Gelman-Rubin diagnostic, effective sample size and autocorrelation time of the RE-SkN model estimated using Bayesian method for simulated dataset} 
  \label{table:convergence}
  \resizebox{0.75\textwidth}{!}{
    \begin{tabular}{lrrrrrr}
    \toprule
    \multicolumn{1}{c}{\multirow{2}[4]{*}{\textbf{Parameters}}} & \multicolumn{3}{c}{\textbf{1 Block}} & \multicolumn{3}{c}{\textbf{4 Blocks}} \\
\cmidrule{2-7}          & \multicolumn{1}{c}{\textbf{$\hat{R}$}} & \multicolumn{1}{c}{$n_{\text{eff}}$} & \multicolumn{1}{c}{\textbf{$\hat{t}$}} & \multicolumn{1}{c}{\textbf{$\hat{R}$}} & \multicolumn{1}{c}{$n_{\text{eff}}$} & \multicolumn{1}{c}{\textbf{$\hat{t}$}} \\
    \midrule
    $\mu$    & \textbf{1.0103} & 1289.4 & \textbf{34.3} & 1.0104 & \textbf{1324.3} & 36.2 \\
    $\omega$ & 1.0194 & 1224.9 & 38.6  & \textbf{1.0053} & \textbf{1404.7} & \textbf{34.1} \\
    $\beta$  & 1.0115 & 1415.3 & 33.8  & \textbf{1.0043} & \textbf{1456.3} & \textbf{33.0} \\
    $\gamma$ & 1.0124 & 1014.8 & 38.0  & \textbf{1.0052} & \textbf{1567.7} & \textbf{30.9} \\
    $\tau_1$ & \textbf{1.0080} & \textbf{1468.4} & \textbf{31.3} & 1.0117 & 1424.6 & 33.5 \\
    $\tau_2$ & 1.0399 & 275.9 & 70.2  & \textbf{1.0074} & \textbf{1579.2} & \textbf{30.6} \\
    $\xi$    & 1.0316 & 565.9 & 83.1  & \textbf{1.0062} & \textbf{1380.2} & \textbf{35.2} \\
    $\varphi$ & 1.0299 & 509.4 & 88.2  & \textbf{1.0075} & \textbf{1373.3} & \textbf{35.3} \\
    $\delta_1$ & 1.0082 & \textbf{1609.9} & \textbf{29.6} & \textbf{1.0052} & 1213.2 & 33.4 \\
    $\delta_2$ & 1.0761 & 111.0 & 436.2 & \textbf{1.0080} & \textbf{1450.8} & \textbf{33.2} \\
    $\sigma^2_u$ & 1.0094 & \textbf{1551.6} & \textbf{27.9} & \textbf{1.0047} & 1366.5 & 35.5 \\
    $\nu$    & 1.3026 & 37.2  & 1341.3 & \textbf{1.0097} & \textbf{1297.1} & \textbf{37.6} \\
    $\lambda$ & 1.0165 & 901.9 & 33.2  & \textbf{1.0095} & \textbf{1461.1} & \textbf{32.5} \\

    \bottomrule
    \end{tabular}% 
}\\
\vspace{0.3cm}
% Result from 10 chains
\end{center}
  \label{tab:addlabel}%
\end{table}%

Second, the estimation results of the proposed RE-SKN model based on the ML method and the MCMC method using the 4 parameter block setting are summarized in Table \ref{table:Sim_Result_1RealMeas}. The optimal method is printed in bold. For each dataset, the true one step-ahead VaR and ES forecasts are calculated and averaged over the 1000 data sets. The average is given in the ‘True’ column. Both the ML estimators and MCMC estimators generate accurate estimates for the parameters and the VaR and ES forecasts, being very close to the true values. However, the MCMC method produce more favorable estimators, both in terms of unbiasedness and precision. All parameters and tail risk forecasts under the MCMC framework, except for $\nu$ parameter, have a smaller bias than those under the ML method. Regarding the estimation precision, all MCMC estimators have smaller root-mean-square errors (RMSE) than their respective ML counterpart.

\begin{table}[H] 
    \captionsetup{font=normalsize, skip=0pt, justification= centering}
    \caption{Summary statistics of the proposed RE-SkN model estimators based on the Maximum Likelihood and MCMC Methods}
    \label{table:Sim_Result_1RealMeas} 
    \begin{center}
  \resizebox{0.7\textwidth}{!}{
    \begin{tabular}{lccccc}
    \toprule
    \textbf{n=2000} & \multicolumn{1}{c}{\multirow{1}[4]{*}{\textbf{True }}} & \multicolumn{2}{c}{\textbf{ML}} & \multicolumn{2}{c}{\textbf{MCMC}} \\
\cmidrule{3-6}    \textbf{Parameter} &       & \multicolumn{1}{c}{\textbf{Mean}} & \multicolumn{1}{c}{\textbf{RMSE}} & \multicolumn{1}{c}{\textbf{Mean}} & \multicolumn{1}{c}{\textbf{RMSE}} \\
    \midrule
    $\mu$    & 0.00     & 0.0028 & 0.0176 & \textbf{0.0001} & \textbf{0.0002} \\
    $\omega$ & -0.12 & -0.1407 & 0.1208 & \textbf{-0.1342} & \textbf{0.0329} \\
    $\beta$  & 0.98  & 0.9579 & 0.1114 & \textbf{0.9777} & \textbf{0.0053} \\
    $\gamma$ & 0.47  & 0.4951 & 0.1658 & \textbf{0.4815} & \textbf{0.0531} \\
    $\tau_1$ & -0.12 & -0.1133 & 0.0665 & \textbf{-0.1210} & \textbf{0.0125} \\
    $\tau_2$ & 0.04  & 0.0512 & 0.0606 & \textbf{0.0407} & \textbf{0.0042} \\
    $\xi$    & -0.17 & -0.0896 & 0.6982 & \textbf{-0.2002} & \textbf{0.1884} \\
    $\varphi$ & 0.93  & 0.9452 & 0.1742 & \textbf{0.9263} & \textbf{0.0267} \\
    $\delta_1$ & -0.09 & -0.0826 & 0.0587 & \textbf{-0.0901} & \textbf{0.0115} \\
    $\delta_2$ & 0.06  & 0.0786 & 0.1266 & \textbf{0.0608} & \textbf{0.0050} \\
    $\sigma_u^2$ & 0.15  & 0.2154 & 0.4433 & \textbf{0.1542} & \textbf{0.0386} \\
    $\nu$    & 4.4   & \textbf{4.3858} & 0.4132 & 4.5075 & \textbf{0.3685} \\
    $\lambda$ & 0.5   & 0.4895 & 0.0962 & \textbf{0.5014} & \textbf{0.0256} \\
    2.5\% $\text{VaR}_{t+1}$ & -0.0763 & -0.0857 & 0.1169 & \textbf{-0.0768} & \textbf{0.0037} \\
    2.5\% $\text{ES}_{t+1}$ & -0.0948 & -0.1062 & 0.1459 & \textbf{-0.0952} & \textbf{0.0055} \\
    1\% $\text{VaR}_{t+1}$ & -0.0916 & -0.1047 & 0.1472 & \textbf{-0.0921} & \textbf{0.0049} \\
    1\% $\text{ES}_{t+1}$ & -0.1130 & -0.1291 & 0.1817 & \textbf{-0.1133} & \textbf{0.0073} \\
    \bottomrule
    \end{tabular}%
}
    \end{center}
%}
  \label{tab:addlabel}%
\end{table}%

\begin{figure}[H]
    \begin{center}
        \captionsetup{font=normalsize, skip=0pt, justification= centering}
        \caption{Histogram of 1000 2.5\% and 1\% VaR true and forecasts: MCMC and ML estimation}
        \label{fig:VaR_histogram}
        \vspace{-3pt}
        \resizebox{0.95\textwidth}{!}{        
        \includegraphics[width=\linewidth]{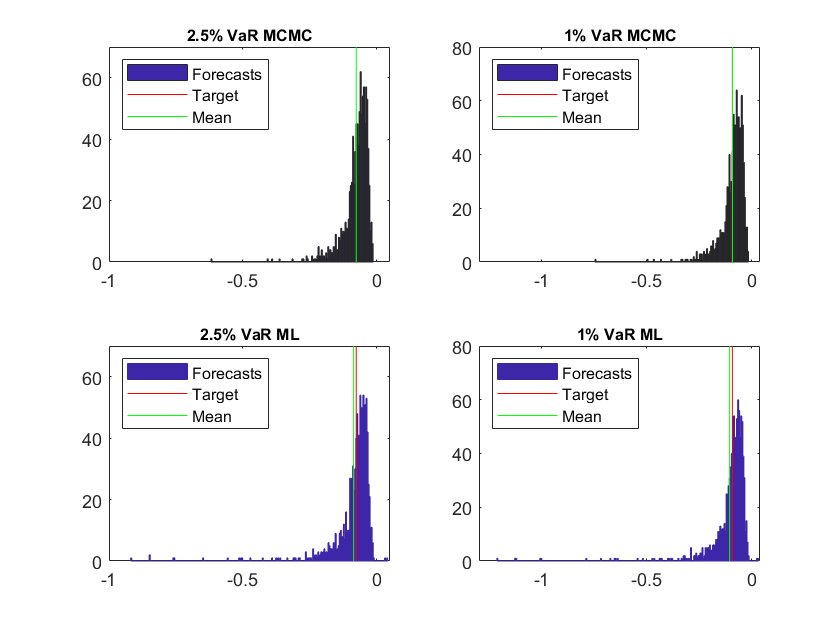}
        }        
    \end{center}
\end{figure}

\begin{figure}[H]
    \begin{center}
        \captionsetup{font=normalsize, skip=0pt, justification= centering}
        \caption{Histogram of 1000 2.5\% and 1\% ES true and forecasts: MCMC and ML estimation}
        \label{fig:ES_histogram}
        \vspace{-3pt}
        \resizebox{0.95\textwidth}{!}{                \includegraphics[width=\linewidth]{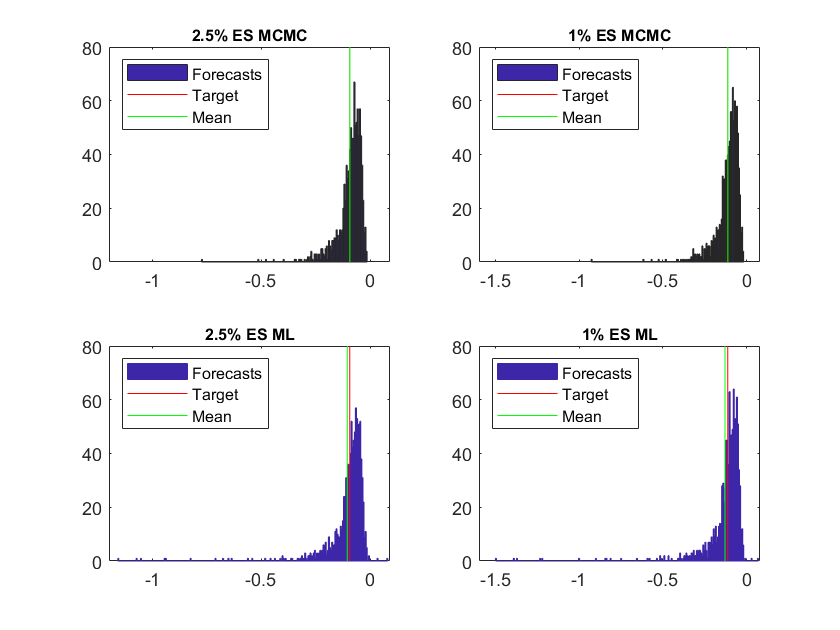}
        }
    \end{center}
\end{figure}

The performance of the MCMC and ML estimators in terms of the unbiasedness and precision of the risk forecasts is also illustrated by using the distribution of 1000 2.5\% and 1\% VaR and ES forecasts in Figure \ref{fig:VaR_histogram} and Figure \ref{fig:ES_histogram}, respectively. Although both the MCMC and ML methods generate VaR and ES forecasts relatively closed to the true value of VaR and ES (represented by the red horizontal lines), it can be seen that the ML method produces more outlying forecasts than the MCMC method. The mean values of both VaR and ES forecasts generated using the MCMC method are also very close to their respective true values, compared to those of the ML method. This might be due to the convergence issue in the ML method, which results in a higher RMSE of the risk forecasts. This suggests that the MCMC method is more preferable than the ML method since it produces more accurate risk forecasts.     


\section{Tail risk forecasting method and evaluation}
\label{sec:forecast}
The realized EGARCH models are estimated using the proposed MCMC procedures with a fixed size in-sample data combined with a rolling window approach to produce 1000 one-step ahead forecasts. In a Bayesian setting, each MCMC iterates of each parameter is used to form MCMC iterates for VaR and ES forecasts at 2.5\% and 1\% quantile levels. The posterior mean estimates of the VaR and ES are then used as the final forecast. 

\subsection{Value-at-risk forecast}
\label{subsec:VaR_forecast}
The original realized EGARCH model estimated using the proposed MCMC method (RE-NN) is used to calculate VaR forecasts by using the inverse cumulative distribution function of the distribution (CDF), $\Phi^{-1}$, as a benchmark to the proposed realized EGARCH model using the standardized Student-\textit{t} distribution (RE-tN) and the standardized skewed Student-\textit{t} distribution (RE-SkN) for the return equation errors.
\begin{eqnarray} \label{VaR_normal}
VaR_\alpha = \mu + \sqrt{h_{t+1}}\Phi^{-1}(\alpha)
\end{eqnarray}

For the standardized Student-\textit{t} distribution, the VaR forecasts are calculated by using the inverse CDF, $t_{\nu}^{-1}$, given by:
\begin{eqnarray}\label{VaR_Student_t}
VaR_{\alpha} = \mu +\sqrt{h_{t+1}} t_v^{-1}(\alpha)\sqrt{\frac{v-2}{v}},
\end{eqnarray}

while for the skewed Student-\textit{t} distribution, the VaR estimation is conducted using the inverse CDF, $skt_{\nu, \lambda}^{-1}$ as specified in \citet{ContinoGerlach2017} as follows: 
\begin{eqnarray} \label{VaR_SkewedStudent_t}
    VaR_{\alpha}=\mu +\sqrt{h_{t+1}}skt^{-1}_{v,\lambda},
\end{eqnarray}
where $skt_{v,\lambda}^{-1}$ is
 \[ skt^{-1}(\alpha \mid v, \lambda)=\left\{ \begin{array} {ll} 
    \frac{1-\lambda}{b}\sqrt{\frac{v-2}{v}}t_{v}^{-1} \Big(\frac{\alpha}{1-\lambda},v\Big)-\frac{a}{b}, & \mbox {if $\alpha < \frac{1-\lambda}{2}$}  \cr
    \frac{1+\lambda}{b}\sqrt{\frac{v-2}{v}}t_{v}^{-1} \Big(0.5+\frac{\alpha}{1+\lambda} \Big(\alpha-\frac{1-\lambda}{2}\Big),v\Big)-\frac{a}{b}, & \mbox {if $\alpha \geq \frac{1-\lambda}{2}$}.
    \end{array} 
    \right.\]

\subsection{Expected shortfall forecasts}
\label{subsec:ES_forecast}
Similar to the VaR, three types of ES forecasts are also calculated. For a normally distributed loss, \citet{McNeilEtAl2005} formulated ES as follows:
\begin{eqnarray}\label{ES_normal}
ES_{\alpha}= \mu + \sqrt{h_{t+1}} \frac{\phi(\Phi^{-1}(\alpha))}{1-\alpha}, 
\end{eqnarray}
where $\phi$ denotes the density of the standard normal distribution and $\Phi^{-1}$ is inverse of the normal CDF.

For the standardized Student $t$-distribution, the ES takes the following form \citep{McNeilEtAl2005}:
\begin{eqnarray}\label{ES_StudentT}
ES_{\alpha}= \mu + \sqrt{h_{t+1}} \frac{g_v(t^{-1}_{v}(\alpha))}{1-\alpha} \Bigg(\frac{v+(t^{-1}_{v}(\alpha))^2}{v-1}\Bigg) \sqrt{\frac{v-2}{v}},
\end{eqnarray}
where $t_{v}$ is the density function, $g_v$ is the density of the standard Student-$t$, $t_v^{-1}$ is the inverse CDF, and $v$ is the estimated degrees of freedom with a restriction $v>2$. 

For the standardized skewed Student-\textit{t} distribution, \citet{ContinoGerlach2017} derived the ES, given the standardized VaR value $\Theta$, as follows: 
\begin{eqnarray}
    ES_{\alpha}=\mu + \sqrt{h_{t+1}} E[\epsilon \mid \epsilon < \Theta],
\end{eqnarray}
\noindent where
\begin{align}
    E[\epsilon \mid < \Theta] &= \frac{1}{skt_{v,\lambda}(\Theta)} \int_{-\infty}^{\Theta}bc\Bigg(1+\frac{1}{v-2}\Bigg(\frac{b\epsilon+a}{1-\lambda}\Bigg)^2\Bigg)^{-\frac{v+1}{2}} \epsilon d \epsilon \cr
    &= \frac{c(1-\lambda)^2}{b \cdot skt_{v,\lambda}(\Theta)} \frac{v-2}{1-v} \Bigg(1+\frac{1}{v-2}\Bigg(\frac{b\Theta +a}{1-\lambda}\Bigg)^2\Bigg)^{\frac{1-v}{2}}-\frac{a}{b},
\end{align}

\noindent where $skt_{v,\lambda}$ is the CDF of the Skewed Student-$t$ distribution:
\[ skt(\epsilon \mid v, \lambda)=\left\{ \begin{array} {ll} 
    (1-\lambda) t_{v} \Big(\sqrt{\frac{v}{v-2}}\Big(\frac{b\epsilon+a}{1-\lambda}\Big),v \Big) & \mbox {if $\epsilon < -\frac{a}{b}$}  \cr
    \frac{1-\lambda}{2} (1+\lambda) \Big[t_{v}\Big(\sqrt{\frac{v}{v-2}}\Big(\frac{b\epsilon+a}{1-\lambda}\Big),v\Big)-0.5\Big] & \mbox {if $\epsilon \geq -\frac{a}{b}$}.
    \end{array} 
    \right.\]

\subsection{Model Evaluation}
Several tests are conducted to assess the accuracy of VaR and ES forecasts. As an initial assessment of VaR forecasting accuracy, the VaR violation rate (VRate) is examined.  VRate is calculated as the proportion of returns exceeding the forecasted VaR level in the forecasting period, as given by:
\begin{eqnarray} \label{VRate}
VRate = \frac{1}{m} \sum_{t=n+1}^{n+m}I(r_t < VaR_t),
\end{eqnarray}
where $n$ is the in-sample size and $m$ is the forecasting sample size.

To formally evaluate the VaR forecast accuracy, we employ three standard VaR backtest procedures that typically focus on coverage tests. These are based on the comparison between the number of times that losses exceed the VaR and the expected number using statistical tests, such the unconditional coverage (UC) test of \citet{Kupiec1995}, the conditional coverage (CC) test of \citet{Christoffersen1998}, and the dynamic quantile (DQ) test of \citet{EngleManganelli2004}. The UC test is employed to test whether the VaR violation rate is statistically different from the confidence level $\alpha$. The CC test improves the UC test by not only testing the joint assumption of the unconditional coverage, but also testing whether the probability of VaR violation is independent over time. The DQ test is a regression-based model of hits variable on a constant, the lagged values of the hit variable, and any function of the past information. In this study, we use 4 lags of the hit variable as recommended by \citet{EngleManganelli2004}. For details regarding those tests, please refer to \citet{Kupiec1995}, \citet{Christoffersen1998}, and \citet{EngleManganelli2004}.

Further, to assess whether the ES forecast is correctly specified, two recently introduced ES backtests are employed. The first ES backtest utilized in this study is
the bivariate Expected Shortfall Regression (ESR) backtest developed by \citet{BayerDimitriadis2018}. The ESR backtest is based on a joint regression model for the quantile and the ES, where the returns $r_t$ is regressed on the ES forecast $\hat{e}_t$ and an intercept:
\begin{eqnarray} \label{bivariate_ESR}
r_t = \alpha+ \beta\hat{e}_t + u^e_t,
\end{eqnarray}
where $ES_\tau (u^e_t|\mathcal{F}_{t-1})=0$.

The null hypothesis of this test is $H_0: (\alpha, \beta) = (0,1)$, against the alternative hypothesis $H_1: (\alpha, \beta) \neq (0,1)$.  Under the null hypothesis, the ES forecast are correctly specified, since this implies that $\hat{e}_t = ES_{\tau}(r_t|\mathcal{F}_{t-1})$. For details on ESR backtest, see \citet{BayerDimitriadis2018}. The ESR backtest is conducted based on the R code downloaded from 'https://rdrr.io/github/BayerSe/esback/'. 

The second ES backtest employed is the test developed by \citet{CouperierLeymarie2019}. This method is based on the theory of multi-quantile regression that exploits the relationship between the VaR and ES by using the definition of a Riemann sum, where the $\alpha$-level ES is approximated as a finite sum of several VaR series as given by:
\begin{eqnarray} \label{ES_Riemann}
ES_{\alpha} \approx \frac{1}{p} \sum_{j=1}^{p} VaR_t(u_j),
\end{eqnarray}
where the risk level $u_j$ is defined by $u_j = \alpha + (j-1)\frac{1-\alpha}{p}$ for $j=1, 2, ..., p$, and $p$ represents the number of quantiles applied in the approximation.

\citet{CouperierLeymarie2019} generalized the idea of \citet{GaglianoneEtAl2011} who introduced the VaR as a regressor in a quantile regression model into the assessment of multiple VaRs. This is conducted by estimating a multi-quantile regression model where  the ex-post losses $\{L_t, t=1,..., T\}$ is regressed on the $p$ VaR forecasts $\{\text{VaR}_t(u_j), t=1,..,T\}_{j=1,2,..,p}$  
\begin{eqnarray} \label{ES_multiquantile_reg}
L_t = \beta_0 (u_j) + \beta_1 (u_j)VaR(u_j) + \epsilon_{j,t} \quad \forall j = 1, 2, ..., p,
\end{eqnarray}
where $\beta_0(u_j)$ and $\beta_1(u_j)$ denote the intercept and the slope parameters at level $u_j$, respectively, and $\epsilon_{j,t}$ denotes the error term at risk level $u_j$ and time $t$ such that the $u_j$-th conditional quantile of $\epsilon_{j,t}$ satisfies $Q_{\epsilon j,t} (u_j;\mathcal{F}_{t-1})=0$. 

Given the multi-quantile regression model in equation (\ref{ES_multiquantile_reg}), the $u_j$-th conditional quantile of $L_t$ is defined as:
\begin{eqnarray}
Q_{L_t}(u_j;\mathcal{F}_{t-1}) = \beta_0 (u_j) + \beta_1(u_j) VaR(u_j) \quad \forall j = 1, 2, ..., p.
\end{eqnarray}

The procedure verifies whether $\text{VaR}(u_j)$ perfectly matches $Q_{L_t}(u_j;\mathcal{F}_{t-1})$. Under the null hypothesis that a sequence of VaR is valid, the parameters satisfy $\beta_0(u_j)=0$ and $\beta_1(u_j)=1$, for $j=1,2,..., p$. These parameter values will produce valid tail distribution of the ES model and ES forecasts. There are four null hypotheses of this ES backtests, which assess different settings that the regression coefficients should fulfill for valid ES forecasts. These hypotheses, $H_{0,J_1}, H_{0,J_2}, H_{0,I}, H_{0,S}$ are given by:
\begin{align}
H_{0,J_1}:& \sum_{j=1}^{p} \big(\beta_0(u_j) + \beta_1(u_j)\big) = p,\\
H_{0,J_2}:& \sum_{j=1}^{p} \beta_0(u_j)=0, \text{ and } \sum_{j=1}^{p} \beta_1(u_j)=p,\\
H_{0,I}:&  \sum_{j=1}^{p} \beta_0(u_j)=0,\\
H_{0,S}:& \sum_{j=1}^{p} \beta_1(u_j)=p,
\end{align}
where $J_1$ and $J_2$ denotes the joint backtests, $I$ refers to the intercept backtest, and $S$ indicates the slope backtest. For details on the ES backtest based on multi-quantile regression (referred hereafter as ES-MQR), please see \citet{CouperierLeymarie2019}. The ES-MQR tests in this paper adapts the companion Matlab code provided by \citet{CouperierLeymarie2019}, available at 'http://www.runmycode.org/companion/view/3358'.

In addition to testing the VaR violations, measuring the size of the uncovered losses is also of paramount importance. To calculate the uncovered losses, \citet{Lopez1999} proposes the use of a loss function, which is based on the distance between the observed returns and the VaR forecasts in lieu of hypothesis testing approach based on statistical tests. The distance represents the losses that have not been covered. One of the most commonly used loss functions is the tick loss function of \citet{GiacominiKomunjer2005} or often referred as quantile score.  As explained by \citet{Gneiting2011}, quantiles are elicitable since the scoring function is strictly consistent. Therefore, the forecast accuracy of VaR can be compared using the following quantile loss function:
\begin{eqnarray} \label{quantile_loss_fn}
    S_\alpha(Q_t,r_t) = \sum_{t=n+1}^{n+m}(r_t-Q_t)(\alpha-I(r_t < Q_t))
\end{eqnarray}
where $Q_{t+1},...,Q_{n+m}$ is a series of quantile forecasts. 
True $\text{VaR}_{\alpha,t}$ minimizes $E \big[ S_{\alpha} ( \text{VaR}_{\alpha,t},r_t) \mid \mathcal{F}_{t-1}\ \big]$. Therefore, the most accurate VaR forecasting model should minimize the scoring function in equation (\ref{quantile_loss_fn}).

The ES, however, is not an elicitable risk measure \citep{Gneiting2011}. Therefore, in general there is no loss function minimized by the true ES. Nevertheless, \citet{FisslerZiegel2016} showed that the ES is elicitable of higher order in the sense that the VaR and ES are jointly elicitable. They developed a strictly consistent function that is minimized by the true VaR and ES:
\begin{align} \label{VaRES_jointloss_Fizzler}
    S_{t}(r_{t}, VaR_{t}, ES_{t})&=         (I_{t}-\alpha)G_{1}(VaR_{t})-I_{t}G_{1}(r_{t}) \cr 
    & +G_{2}(ES_{t})\Big(ES_{t}-VaR_{t}+\frac{I_{t}}{\alpha}(VaR_{t}-r_{t})\Big)\cr
        &-H(ES_{t})+a(r_{t}),
\end{align}
where $I_{t}=1$ if $r_{t} < VaR_{t}$ and 0 otherwise for $t =1,...,T$, $G_{1}()$ is increasing, $G_{2}()$ is strictly increasing and strictly convex, $G_{2}=H'$ and $\text{lim}_{x \rightarrow{- \infty}} G_{2}(x)=0$ and a($\cdot$) is a real valued integrable function. 

This study employs two special joint loss functions. The first one is based on the recommendation of \citet{FisslerZiegel2016} which has been used in the study of \citet{WangEtAl2019}, while the second one is based on \citet{Taylor2019}. The first joint loss function considers $G_1(x) = x, G_2(x) = \exp(x), H(x) =  \exp(x)$ and $a(r_t) = 1-\log(1-\alpha)$ that gives the following scoring function:
\begin{align} \label{VaRES_jointloss_specific}
    S_{t}(r_{t}, VaR_{t}, ES_{t})&=         (I_{t}-\alpha)(VaR_{t})-I_{t}(r_{t})\cr
        &+\text{exp}(ES_{t})\Big(ES_{t}-VaR_{t}+\frac{I_{t}}{\alpha}(VaR_{t}-r_{t})\Big)\cr
        &-\text{exp}(ES_{t})+1-\log(1-\alpha),
\end{align}
where the loss function $S=\sum_{t-1}^{T}S_t$ is strictly consistent and jointly minimized by the true VaR and ES series.

The second  joint loss function is a method for predicting the ES corresponding to the VaR forecasts by using the equivalence between quantile regression and maximum likelihood based on an asymmetric Laplace (AL) density \citep{Taylor2019}. In this framework, a joint model of conditional VaR and ES is estimated by maximizing the AL log-likelihood. The scoring function proposed by \citet{Taylor2019} is based on the scoring function of \citet{FisslerZiegel2016} in equation (\ref{VaRES_jointloss_Fizzler}) by considering $G_1 = 0, G_2 = - 1/x, H(x) =  - \log (-x)$ and $ a(r_t) = 1-\log(1-\alpha)$. The function is the negative of the AL log-likelihood, referred to as the AL log score, which takes the following form:
\begin{eqnarray} \label{AL_log_score}
    S(Q_{t},ES_{t},y_{t})=-\log\Bigg(\frac{\alpha-1}{ES_{t}}\Bigg)-\frac{(y_{t}-Q_{t})(\alpha-I(y_{t} \leq Q_{t}))}{\alpha ES_{t}}.
\end{eqnarray}

A strictly consistent scoring function can be used as the loss function in model comparison \citep{GneitingRaftery2007}. The scoring function in equation (\ref{AL_log_score}) is strictly consistent and the score average across a sample provides a joint measure of VaR and ES forecast accuracy.

In the next step, the AL log score is then used as the loss function in the calculation of the Model Confidence Set (MCS). The MCS, developed by \citet{HansenEtAl2011}, consists of a sequence of tests used to construct a set of “superior” models (SSM) for which the null hypothesis of equal predictive ability (EPA) is not rejected. The MCS procedure will produce a set of models with a given confidence level.  The MCS test is conducted using two methods, i.e. R method and SQ method. The first uses the sums of absolute values in the calculation of MCS test statistic, while the latter uses the summed squares. For details, see \citet{HansenEtAl2011} page 465. The MCS approach is available in the Matlab MFE toolbox of Kevin Sheppard and the associated Matlab code can be downloaded from 'https://www.kevinsheppard.com/code/matlab/mfe-toolbox/'. 

\section{Empirical Study}
\label{sec:empirical}
In this section, we present the empirical results of the proposed models, the RE-tN and RE-SkN type models, estimated using the MCMC approach using returns and 3 types of realized measures (individual or combined) specified in Section \ref{sec:realized_measures}. The risk forecast performance of the models are evaluated based on various tests discussed in Section \ref{sec:forecast} and compared with the original realized EGARCH model estimated also within a Bayesian setting (the RE-NN type models) . Several GARCH-family type models, all with Student-\textit{t} distribution, are also included for comparison purposes, such as the GARCH, EGARCH, and GJR-GARCH models. We also include a GARCH-t-EVT model to produce conditional quantile and ES estimates by applying the peaks over threshold of extreme value theory (EVT) method to the standardized residuals of the GARCH model with Student-\textit{t} distribution. We follow \citet{Taylor2008} and set the threshold as the 10\% unconditional quantile for the lower tail. For details regarding GARCH-t-EVT models, see \citet{McNeilFrey2000} and Section 7.2.3 in \citet{McNeilEtAl2005}. The comparison models also include a filtered GARCH approach (GARCH-t-HS) used in \citet{WangEtAl2019}, where a standardized VaR and ES are estimated based on a GARCH with with Student-\textit{t} distribution (GARCH-t). The returns ($r_1, .., r_n$) are standardized by using by their estimated conditional standard deviation, $rt/\sqrt{\hat{h}_t}$. The standardized VaR and ES are then multiplied by the forecast $\sqrt{\hat{h}_{n+1}}$ from the GARCH-t model. We also generate VaR and ES forecasts based on symmetric absolute value conditional autoregressive expectiles (CARE-SAV) model of \citet{Taylor2008} and the realized GARCH models of \citet{HansenEtAl2011} with a Student-\textit{t} distribution for the observation equation errors for the return (RG-tN). The GARCH-type models are estimated using the Econometrics toolbox in Matlab, while the GARCH-t-EVT, CARE-SAV, realized GARCH, and realized EGARCH models are estimated using Matlab code developed by authors. The parameters of the models are used to generate 1000 one-step-ahead VaR and ES forecasts.
 
\subsection{Data}

This study utilizes the data of daily, including open, high, low and closing prices, as well as high frequency data, observed at 5-minute and 1-minute intervals within trading hours. The data are downloaded from Thomson Reuters Tick History. The data are collected for 7 market indices: ASX200 (Australia), DAX (Germany), FTSE 100 (UK), Hang Seng (Hong Kong), NASDAQ (US), SMI (Swiss) and S\&P500 (US), covering the period of January 2000 to June 2016.

The daily return is calculated based on the close-to-close prices, including overnight price movements. The sub-sampled RV (RVSS) and sub-sampled RR (RRSS) used in this study are based on the dataset used in \citet{WangEtAl2019}, in which the daily RV and RR are calculated based on the 5-minute data, and the 1-minute data are used to generate daily RVSS and RRSS. Meanwhile, the data for the realized kernel using a Parzen weight function is obtained from the Oxford-Man Institute's realized library, Library Version: 0.3, which also originates from Thomson Reuters Tick History. The realized measures might be biased downward measure for the true volatility, since they are calculated during the market is open, ignoring the overnight price movement. How the realized EGARCH models adjust for such downward bias is discussed in section \ref{sec:param_estimates}. 

\subsection{Model Parameter Estimates}
\label{sec:param_estimates}
Before discussing tail risk forecast evaluations, the estimated parameters for all forecasting steps of the realized EGARCH models are presented. In this section, we take an example of the estimated parameters for the S\&P500 dataset presented in Table \ref{table:parameter_SP-1} and Table \ref{table:parameter_SP-2}. In general, the parameter estimates are in line with with those in Table 3 of \citet{HansenHuang2016}. First, $\gamma$ parameter, which provides the channel for the impact of the realized measures on future volatility, ranges from 0.141 to 0.357 in the models with single realized measure. Furthermore, estimated $\gamma$ in the models using the RVSS and RRSS are much greater (0.237 and 0.222 and on average, respectively) than those using the RK (0.046 on average), indicating that the RVSS and RRSS provide a stronger signal than the RK. The estimated $\gamma$ in models employing the RVSS and RRSS is several times larger than the typical value for $\alpha$ (about 0.05) in a conventional GARCH model, which measures the coefficient associated with squared returns. This suggests that the realized measures offers a stronger signal about future volatility than the squared return does. 

Second, the estimates of $\beta$ are close to 1 in all cases and $\beta - \gamma'\varphi_k$ are also relatively high although they are less than unity, suggesting volatility persistence. The volatility is less persistent in the models employing the RVSS and RRSS, compared to those employing the RK. For example, the average estimates of $\beta - \gamma'\varphi_k$ of the RE-RV-SkN and RE-RR-SkN models are 0.648 and 0.622, respectively, while that of the RE-RK-SkN model is 0.829.   

Third, the leverage functions ($\tau_1, \tau_2, \delta_1, \delta_2$) are of the expected sign: $\tau_1$ and $\delta_1$ are negative and $\tau_2$ and $\delta_2$ positive. The degree of asymmetry in the leverage effect is consistent across models, where the average estimate of $\tau_1$ is -0.176 and $\tau_2$ 0.037. However, the estimated $\delta_1$ and $\delta_2$ tend to be smaller (more negative for $\delta_1$) when using the RVSS and RRSS than the RK. The fact that $\delta_1 < 0$ indicates that  $x_t$ will be larger when $\epsilon_t < 0$ than when $\epsilon_t > 0$, which will make $h_{t+1}$ larger when $\epsilon_t < 0$ through variance equation if $\gamma > 0$. This result of leverage effects is in line with the well-known stock market phenomenon that there is a negative correlation between today's return and tomorrow's volatility.

\newpage
\begin{table}[H] 
\begin{center}
\captionsetup{font=normalsize, skip=0pt, justification= centering}
\caption{Estimated parameter mean for all forecasting steps for each parameter: S\&P500-Part 1} \label{table:parameter_SP-1}
  \resizebox{1.15\textwidth}{!}{
    \begin{tabular}{lcccccccccccccc}
    \toprule
    \textbf{Model} & \textbf{$\mu$} & \textbf{$\omega$} & \textbf{$\beta$} & \textbf{$\gamma_1$} & \textbf{$\gamma_2$} & \textbf{$\gamma_3$} & \textbf{$\tau_1$} & \textbf{$\tau_2$} & \textbf{$\xi_1$} & \textbf{$\xi_2$} & \textbf{$\xi_3$} & \textbf{$\varphi_1$} & \textbf{$\varphi_2$} & \textbf{$\varphi_3$} \\
    \midrule
    RE-RV-NN & 0.0001 & -0.322 & 0.965 & 0.319 &       &       & -0.179 & 0.037 & -0.934 &       &       & 0.967 &       &  \\
    RE-RV-SkN & 0.0000 & -0.324 & 0.965 & 0.336 &       &       & -0.186 & 0.039 & -1.165 &       &       & 0.943 &       &  \\
    RE-RV-tN & 0.0003 & -0.305 & 0.968 & 0.335 &       &       & -0.183 & 0.038 & -1.158 &       &       & 0.944 &       &  \\
    RE-RR-NN & 0.0000 & -0.326 & 0.965 &       & 0.344 &       & -0.177 & 0.039 &       & -1.066 &       &       & 0.977 &  \\
    RE-RR-SkN & 0.0000 & -0.327 & 0.965 &       & 0.357 &       & -0.183 & 0.041 &       & -1.227 &       &       & 0.961 &  \\
    RE-RR-tN & 0.0003 & -0.313 & 0.967 &       & 0.356 &       & -0.179 & 0.040 &       & -1.210 &       &       & 0.963 &  \\
    RE-RK-NN & 0.0001 & -0.264 & 0.972 &       &       & 0.141 & -0.174 & 0.031 &       &       & -0.538 &       &       & 1.003 \\
    RE-RK-SkN & 0.0001 & -0.272 & 0.971 &       &       & 0.144 & -0.181 & 0.032 &       &       & -0.694 &       &       & 0.987 \\
    RE-RK-tN & 0.0003 & -0.254 & 0.973 &       &       & 0.144 & -0.178 & 0.032 &       &       & -0.662 &       &       & 0.990 \\
    RE-RV-RR-NN & 0.0001 & -0.355 & 0.962 & 0.185 & 0.125 &       & -0.162 & 0.042 & -0.239 & -0.354 &       & 1.041 & 1.054 &  \\
    RE-RV-RR-SkN & 0.0001 & -0.348 & 0.963 & 0.198 & 0.133 &       & -0.172 & 0.044 & -0.707 & -0.817 &       & 0.991 & 1.004 &  \\
    RE-RV-RR-tN & 0.0003 & -0.334 & 0.965 & 0.199 & 0.128 &       & -0.168 & 0.043 & -0.587 & -0.702 &       & 1.005 & 1.018 &  \\
    RE-RV-RK-NN & 0.0002 & -0.330 & 0.965 & 0.256 &       & 0.011 & -0.173 & 0.033 & -0.220 &       & -0.150 & 1.043 &       & 1.043 \\
    RE-RV-RK-SkN & 0.0001 & -0.339 & 0.964 & 0.271 &       & 0.013 & -0.182 & 0.034 & -0.539 &       & -0.422 & 1.009 &       & 1.015 \\
    RE-RV-RK-tN & 0.0004 & -0.318 & 0.967 & 0.266 &       & 0.013 & -0.177 & 0.034 & -0.425 &       & -0.337 & 1.022 &       & 1.024 \\
    RE-RR-RK-NN & 0.0001 & -0.339 & 0.964 &       & 0.282 & 0.011 & -0.174 & 0.034 &       & -0.408 & -0.205 &       & 1.048 & 1.038 \\
    RE-RR-RK-SkN & 0.0001 & -0.347 & 0.963 &       & 0.295 & 0.012 & -0.182 & 0.036 &       & -0.700 & -0.457 &       & 1.016 & 1.011 \\
    RE-RR-RK-tN & 0.0003 & -0.333 & 0.965 &       & 0.291 & 0.013 & -0.178 & 0.036 &       & -0.592 & -0.360 &       & 1.028 & 1.021 \\
    RE-RV-RR-RK-NN & 0.0002 & -0.349 & 0.963 & 0.158 & 0.120 & 0.015 & -0.167 & 0.039 & -0.244 & -0.360 & -0.117 & 1.041 & 1.053 & 1.047 \\
    RE-RV-RR-RK-SkN & 0.0002 & -0.353 & 0.962 & 0.163 & 0.118 & 0.015 & -0.169 & 0.039 & -0.275 & -0.407 & -0.162 & 1.037 & 1.048 & 1.042 \\
    RE-RV-RR-RK-tN & 0.0004 & -0.337 & 0.965 & 0.161 & 0.117 & 0.016 & -0.165 & 0.040 & -0.246 & -0.368 & -0.138 & 1.042 & 1.053 & 1.045 \\
    \midrule
    \textbf{Average} & \textbf{0.0002} & \textbf{-0.323} & \textbf{0.966} & \textbf{0.237} & \textbf{0.222} & \textbf{0.046} & \textbf{-0.176} & \textbf{0.037} & \textbf{-0.562} & \textbf{-0.684} & \textbf{-0.353} & \textbf{1.007} & \textbf{1.019} & \textbf{1.022} \\
    \bottomrule
    \end{tabular}%
}
\end{center}
  \label{tab:addlabel}%
\end{table}%

\begin{table}[H]
  \begin{center}
  \captionsetup{font=normalsize, skip=0pt, justification= centering}
  \caption{Estimated parameter mean for all forecasting steps for each parameter: S\&P500-Part 2} \label{table:parameter_SP-2}
  \resizebox{1.1\textwidth}{!}{
    \begin{tabular}{lcccccccccccc}
    \toprule
    \textbf{Model} & \textbf{$\delta_{11}$} & \textbf{$\delta_{12}$} & \textbf{$\delta_{13}$} & \textbf{$\delta_{21}$} & \textbf{$\delta_{22}$} & \textbf{$\delta_{23}$} & \textbf{$\sigma_{u_1}^2$} & \textbf{$\sigma_{u_2}^2$} & \textbf{$\sigma_{u_3}^2$} & \textbf{$\nu$} & \textbf{$\nu_{skw}$} & \textbf{$\lambda$} \\
    \midrule
    RE-RV-NN & -0.159 &       &       & 0.064 &       &       & 0.194 &       &       &       &       &  \\
    RE-RV-SkN & -0.160 &       &       & 0.065 &       &       & 0.194 &       &       &       & 10.266 & -0.167 \\
    RE-RV-tN & -0.156 &       &       & 0.064 &       &       & 0.194 &       &       & 9.594 &       &  \\
    RE-RR-NN &       & -0.153 &       &       & 0.065 &       &       & 0.175 &       &       &       &  \\
    RE-RR-SkN &       & -0.154 &       &       & 0.066 &       &       & 0.175 &       &       & 10.688 & -0.165 \\
    RE-RR-tN &       & -0.150 &       &       & 0.066 &       &       & 0.175 &       & 10.187 &       &  \\
    RE-RK-NN &       &       & -0.053 &       &       & 0.205 &       &       & 0.413 &       &       &  \\
    RE-RK-SkN &       &       & -0.056 &       &       & 0.204 &       &       & 0.413 &       & 10.599 & -0.166 \\
    RE-RK-tN &       &       & -0.042 &       &       & 0.205 &       &       & 0.413 & 9.750 &       &  \\
    RE-RV-RR-NN & -0.158 & -0.151 &       & 0.063 & 0.066 &       & 0.194 & 0.176 &       &       &       &  \\
    RE-RV-RR-SkN & -0.159 & -0.153 &       & 0.064 & 0.067 &       & 0.195 & 0.176 &       &       & 10.568 & -0.162 \\
    RE-RV-RR-tN & -0.155 & -0.147 &       & 0.064 & 0.067 &       & 0.195 & 0.176 &       & 9.766 &       &  \\
    RE-RV-RK-NN & -0.158 &       & -0.050 & 0.062 &       & 0.202 & 0.194 &       & 0.402 &       &       &  \\
    RE-RV-RK-SkN & -0.159 &       & -0.053 & 0.063 &       & 0.202 & 0.194 &       & 0.402 &       & 10.797 & -0.161 \\
    RE-RV-RK-tN & -0.156 &       & -0.041 & 0.063 &       & 0.203 & 0.194 &       & 0.401 & 10.093 &       &  \\
    RE-RR-RK-NN &       & -0.152 & -0.054 &       & 0.064 & 0.204 &       & 0.176 & 0.402 &       &       &  \\
    RE-RR-RK-SkN &       & -0.153 & -0.056 &       & 0.065 & 0.203 &       & 0.176 & 0.402 &       & 11.078 & -0.160 \\
    RE-RR-RK-tN &       & -0.150 & -0.043 &       & 0.065 & 0.204 &       & 0.176 & 0.401 & 10.370 &       &  \\
    RE-RV-RR-RK-NN & -0.157 & -0.150 & -0.050 & 0.063 & 0.066 & 0.203 & 0.195 & 0.176 & 0.403 &       &       &  \\
    RE-RV-RR-RK-SkN & -0.158 & -0.151 & -0.051 & 0.063 & 0.065 & 0.203 & 0.195 & 0.176 & 0.403 &       & 11.047 & -0.157 \\
    RE-RV-RR-RK-tN & -0.154 & -0.147 & -0.038 & 0.063 & 0.066 & 0.204 & 0.195 & 0.175 & 0.403 & 9.762 &       &  \\
    \midrule
    \textbf{Average} & \textbf{-0.157} & \textbf{-0.151} & \textbf{-0.049} & \textbf{0.064} & \textbf{0.066} & \textbf{0.203} & \textbf{0.194} & \textbf{0.176} & \textbf{0.405} & \textbf{9.932} & \textbf{10.720} & \textbf{-0.162} \\
    \bottomrule
    \end{tabular}%
}
\end{center}
  \label{tab:addlabel}%
\end{table}%

\newpage
Fourth, the realized measures are not perfectly unbiased estimator of the true volatility. For the realized measures to be unbiased, $\xi$ and $\varphi$ would be 0 and 1, respectively. The realized measures are measured from market open to close so they may slightly underestimate volatility on average. Nevertheless, the parameters in the realized EGARCH models can adjust for the bias. The estimated $\varphi < 1$ is in fact close to 1 in all cases, suggesting that the realized measures are roughly proportional to the conditional variance minus a negative correction given by the estimate of $\xi$. % The estimated $\xi$ is negative in all models, suggesting downward biased, but it is lower for models employing RK than those using RVSS and RRSS.

Fifth, parameter $\sigma_u^2$ partially indicates how much error contained in the realized measures, compared with $\log(h)$, but the log-scale makes this difficult to interpret. The models using the RRSS generates a smaller $\sigma_u$ compared to those using the RVSS and RK, meaning that the RRSS can potentially provide extra efficiency. This is in line with the findings of \citet{MartensVanDijk2007}, \citet{ChristensenPodolskij2007}, and \citet{WangEtAl2019} that conclude that the realized range has lower mean squared errors than the realized variance, which may allow models employing the realized range to produce a higher accuracy and efficiency in volatility estimation and forecasting.  
    
Sixth, the estimates of $\nu$ are around 9 to 12, indicating that the choices of the Student-\textit{t} and skewed Student-\textit{t} return equation errors are justified, compared with the usual Gaussian distribution. Lastly, the estimated $\lambda$ is negative, suggesting that the return is skewed to the left.

\subsection{VaR and ES Backtests}
\label{sec:model_evaluation}

A prudent risk management requires accurate risk forecasts. On the one hand, financial regulators are usually concerned about how many times losses exceed the VaR (the VaR violations) and the uncovered loss size. On the other hand, risk managers must maintain the balance between the safety goal and the profit maximization goal. An excessively high VaR will result in an inefficiency since financial institutions face large opportunity costs of capital due to reserving too much capital to buffer againts the market risks. 

The VaR violation rates (VRates) for each model for all market indices are presented in Table \ref{table:VRate25pct} for 2.5\% forecasting and Table \ref{table:VRate1pct} for 1\% forecasting. Overall, the proposed RE-tN and RE-SkN type models generate more optimal VaR forecast series compared to other competing models, except for the CARE-SAV model for the 2.5\% forecasting. The proposed models are consistently conservative, in the sense that they the VRates are closer to the nominal VRates of 2.5\% and 1\%.  Other competing models tend to be anti-conservative, generating VaR forecasts much higher than 2.5\% and 1\%, except for the CARE-SAV and GARCH-t-HS models. The GARCH-t-EVT in particular produces too low VaR forecasts for the 2.5\% level.

For the 2.5\% VaR forecasting, the model with the average VRate closest to the nominal VRate of 2.5\% is the CARE-SAV model (2.54\%), followed by the RE-RR-tN model (2.59\%), RE-RV-RR-tN model (2.60\%), and GARCH-t-HS model (2.60\%). Individually, the CARE-SAV model produces the most accurate VRates for 3 out of 7 indices, while the RE-RV-SkN, RE-RR-tN, and RE-RK-SkN models produce the most accurate VRates for 2 out of 7 indices. 

The favorable performance of the proposed RE-SkN and RE-tN type models is more distinctive for 1\% VaR forecasting level. The best models generating average VRate closest to the nominal VRate of 1\% are the RE-RR-tN and RE-RK-tN models (both 1\%), followed by the RE-RR-RK-tN model (0.99\%) and RE-RV-RR-RK-tN model (1.04\%). Individually, the model that generates the most accurate VRates is the RE-RV-RR-RK-SkN model (4 out of 7 indices), followed by the RE-RR-SkN and RE-RV-RR-RK-tN models (3 out of 7 indices). 

With respect to the number of realized measure employed, there is no clear pattern whether using multiple realized measures improves the 2.5\% VaR forecast accuracy. However, a clear pattern can be observed for the 1\% forecasting level, where models using multiple realized measures produce VRate closer to 1\% more frequently than those using single realized measure. The result also suggests that the models utilizing the RRSS (individually or jointly with other realized measures) with the Student-\textit{t} and Skewed Student-\textit{t} distributions for observation equation errors tend to be more conservative and produce relatively accurate VaR violation rates.

\begin{table}[H]
\begin{center}
  \captionsetup{font=normalsize, skip=0pt, justification= centering}
  \caption{2.5\% VaR forecasting violation rate}
  \label{table:VRate25pct}
    \resizebox{0.95\textwidth}{!}{
    \begin{tabular}{lcccccccc}
    \toprule
    \multicolumn{1}{p{7.955em}}{\textbf{Model}} & \textbf{ASX200} & \textbf{DAX} & \textbf{FTSE} & \textbf{HK} & \textbf{NASDAQ} & \textbf{SMI} & \textbf{S\&P500} & \textbf{Average} \\
    \midrule
    GARCH-t & \textit{\textbf{3.9}} & \textit{\textbf{4.0}} & \textit{\textbf{4.2}} & 3.3   & \textit{\textbf{3.8}} & \textit{\textbf{3.5}} & \textit{\textbf{3.6}} & 3.76 \\
    EGARCH-t & 3.3   & 3.5   & \textit{\textbf{3.7}} & 2.8   & \textit{\textbf{3.7}} & 3.5   & 2.9   & 3.34 \\
    GJR-t & \textit{\textbf{3.8}} & \textit{\textbf{3.6}} & \textit{\textbf{3.6}} & 2.8   & \textit{\textbf{3.8}} & 3.0   & 3.3   & 3.41 \\
    GARCH-t-HS & \textbf{2.6} & \cellcolor[rgb]{ .573,  .816,  .314}\textbf{2.4} & 3.1   & 2.7   & \textbf{2.4} & 2.8   & 2.2   & 2.60 \\
    GARCH-t-EVT & \textit{\textbf{0.3}} & 1.7   & 1.7   & \textit{\textbf{1.3}} & 1.8   & \textit{\textbf{1.4}} & 1.8   & 1.43 \\
    CARE-SAV & \textbf{2.6} & \cellcolor[rgb]{ .573,  .816,  .314}\textbf{2.6} & \cellcolor[rgb]{ .573,  .816,  .314}\textbf{2.6} & 2.8   & \cellcolor[rgb]{ .573,  .816,  .314}\textbf{2.5} & \textbf{2.4} & \textbf{2.3} & 2.54 \\
    RG-RV-tN & 2.8   & \textit{\textbf{4.4}} & \textit{\textbf{4.2}} & 3.4   & \textit{\textbf{3.6}} & \textit{\textbf{3.9}} & \textit{\textbf{4.5}} & 3.83 \\
    RG-RR-tN & \cellcolor[rgb]{ .573,  .816,  .314}\textbf{2.5} & \textit{\textbf{4.0}} & \textit{\textbf{3.8}} & \textbf{2.4} & 2.9   & \textit{\textbf{4.2}} & 3.5   & 3.33 \\
    RG-RK-tN & \textit{\textbf{3.6}} & \textit{\textbf{4.4}} & \textit{\textbf{4.1}} & 3.2   & \textit{\textbf{4.1}} & \textit{\textbf{1.2}} & 3.3   & 3.41 \\
    RE-RV-NN & 3.1   & \textit{\textbf{4.0}} & \textit{\textbf{4.3}} & 2.9   & \textit{\textbf{3.7}} & 3.2   & \textit{\textbf{3.6}} & 3.54 \\
    RE-RV-SkN & \cellcolor[rgb]{ .573,  .816,  .314}\textbf{2.5} & 3.4   & 3.2   & \cellcolor[rgb]{ .573,  .816,  .314}\textbf{2.5} & 2.9   & 2.9   & 3.3   & 2.96 \\
    RE-RV-tN & 2.7   & 3.4   & 3.4   & 2.3   & \cellcolor[rgb]{ .573,  .816,  .314}\textbf{2.5} & \textbf{2.4} & 3.0   & 2.81 \\
    RE-RR-NN & 2.9   & \textit{\textbf{3.9}} & \textit{\textbf{3.8}} & \textbf{2.4} & 3.2   & 3.3   & 3.5   & 3.29 \\
    RE-RR-SkN & 2.1   & 3.2   & 3.0   & 1.8   & \cellcolor[rgb]{ .573,  .816,  .314}\textbf{2.5} & 3.0   & 2.9   & 2.64 \\
    RE-RR-tN & \cellcolor[rgb]{ .573,  .816,  .314}\textbf{2.5} & 3.1   & 3.1   & \textit{\textbf{1.4}} & \cellcolor[rgb]{ .573,  .816,  .314}\textbf{2.5} & 2.8   & \textbf{2.7} & 2.59 \\
    RE-RK-NN & 3.1   & \textit{\textbf{4.1}} & \textit{\textbf{3.9}} & 2.8   & \textit{\textbf{3.6}} & \cellcolor[rgb]{ .573,  .816,  .314}\textbf{2.5} & 3.2   & 3.31 \\
    RE-RK-SkN & 2.8   & 3.3   & 3.1   & \cellcolor[rgb]{ .573,  .816,  .314}\textbf{2.5} & 3.0   & 1.7   & \cellcolor[rgb]{ .573,  .816,  .314}\textbf{2.6} & 2.71 \\
    RE-RK-tN & 1.7   & 3.2   & 2.9   & 1.9   & 3.1   & \textit{\textbf{1.1}} & \cellcolor[rgb]{ .573,  .816,  .314}\textbf{2.4} & 2.33 \\
    RE-RV-RR-NN & 2.7   & \textit{\textbf{4.0}} & \textit{\textbf{4.0}} & \cellcolor[rgb]{ .573,  .816,  .314}\textbf{2.5} & 3.3   & 3.3   & \textit{\textbf{3.7}} & 3.36 \\
    RE-RV-RR-SkN & 1.8   & 3.3   & 3.0   & 2.1   & \textbf{2.6} & 2.9   & 3.1   & 2.69 \\
    RE-RV-RR-tN & 2.2   & 3.2   & 3.2   & \textit{\textbf{1.5}} & \textbf{2.6} & 2.7   & 2.8   & 2.60 \\
    RE-RV-RK-NN & 3.2   & \textit{\textbf{4.1}} & \textit{\textbf{4.3}} & 2.9   & \textit{\textbf{3.7}} & 3.1   & \textit{\textbf{3.9}} & 3.60 \\
    RE-RV-RK-SkN & 2.7   & 4.1   & 3.1   & \cellcolor[rgb]{ .573,  .816,  .314}\textbf{2.5} & 3.0   & 2.8   & 3.2   & 3.06 \\
    RE-RV-RK-tN & 3.0   & 3.4   & 3.2   & 2.1   & 2.8   & \textbf{2.4} & 3.0   & 2.84 \\
    RE-RR-RK-NN & 3.2   & \textit{\textbf{4.0}} & \textit{\textbf{3.8}} & 2.3   & 3.4   & 3.2   & \textit{\textbf{3.6}} & 3.36 \\
    RE-RR-RK-SkN & 2.3   & 3.4   & \textbf{2.8} & 1.9   & \textbf{2.6} & 2.9   & 2.8   & 2.67 \\
    RE-RR-RK-tN & \textbf{2.6} & 3.2   & 2.9   & \textit{\textbf{1.5}} & \cellcolor[rgb]{ .573,  .816,  .314}\textbf{2.5} & 1.2   & \textbf{2.7} & 2.37 \\
    RE-RV-RR-RK-NN & 3.0   & \textit{\textbf{4.1}} & \textit{\textbf{4.1}} & \textbf{2.4} & 3.4   & 3.2   & \textit{\textbf{3.8}} & 3.43 \\
    RE-RV-RR-RK-SkN & 2.1   & 4.1   & 3.1   & 2.0   & \textbf{2.6} & 2.8   & 3.1   & 2.83 \\
    RE-RV-RR-RK-tN & 2.3   & 3.4   & 3.1   & 1.6   & \cellcolor[rgb]{ .573,  .816,  .314}\textbf{2.5} & 2.7   & \textbf{2.7} & 2.61 \\
    \bottomrule
    \multicolumn{9}{p{45em}}{Note: For individual models, green highlight indicates the favoured model, bold indicates the 2nd ranked model, bold italics indicates that the violation rate is significantly different to 2.5\% by the UC test.}\\    
    \end{tabular}%
}
\end{center}
\label{tab:addlabel}%
\end{table}%

    % Table generated by Excel2LaTeX from sheet 'Sheet2'
\begin{table}[H]
  \begin{center}
  \captionsetup{font=normalsize, skip=0pt, justification= centering}
  \caption{1\% VaR forecasting violation rate}
    \label{table:VRate1pct}
    \resizebox{0.95\textwidth}{!}{
    \begin{tabular}{lcccccccc}
    \toprule
    \multicolumn{1}{p{7.955em}}{\textbf{Model}} & \textbf{ASX200} & \textbf{DAX} & \textbf{FTSE} & \textbf{HK} & \textbf{NASDAQ} & \textbf{SMI} & \textbf{S\&P500} & \textbf{Average} \\
    \midrule
    GARCH-t & \textit{\textbf{2.1}} & 1.5   & \textit{\textbf{1.8}} & \textit{\textbf{1.7}} & \textit{\textbf{1.9}} & 1.5   & 1.6   & 1.73 \\
    EGARCH-t & 1.5   & 1.6   & 1.4   & 1.4   & \textit{\textbf{1.7}} & \textit{\textbf{1.9}} & 1.4   & 1.56 \\
    GJR-t & \textit{\textbf{1.8}} & 1.4   & \textit{\textbf{2.0}} & 1.3   & 1.6   & 1.5   & 1.3   & 1.56 \\
    GARCH-t-HS & 1.3   & \cellcolor[rgb]{ .573,  .816,  .314}\textbf{1.1} & 1.3   & \textbf{1.1} & \cellcolor[rgb]{ .573,  .816,  .314}\textbf{0.9} & \textbf{1.1} & \textbf{0.9} & 1.10 \\
    GARCH-t-EVT & \textit{\textbf{0.3}} & \textit{\textbf{1.7}} & \textit{\textbf{1.7}} & 1.3   & \textit{\textbf{1.8}} & 1.4   & 1.6   & 1.40 \\
    CARE-SAV & 0.8   & \cellcolor[rgb]{ .573,  .816,  .314}\textbf{0.9} & 1.2   & 0.8   & \cellcolor[rgb]{ .573,  .816,  .314}\textbf{0.9} & 0.8   & \textbf{1.1} & 0.93 \\
    RG-RV-tN & \textbf{1.1} & \textit{\textbf{2.5}} & \textit{\textbf{2.3}} & \textit{\textbf{1.9}} & \textit{\textbf{2.1}} & \textit{\textbf{2.3}} & \textit{\textbf{2.6}} & 2.11 \\
    RG-RR-tN & \cellcolor[rgb]{ .573,  .816,  .314}\textbf{1.0} & \textit{\textbf{2.2}} & 1.5   & 1.3   & 1.4   & \textit{\textbf{2.5}} & \textit{\textbf{2.0}} & 1.70 \\
    RG-RK-tN & 1.2   & \textit{\textbf{2.6}} & \textit{\textbf{1.8}} & 1.5   & \textit{\textbf{2.6}} & 0.8   & \textit{\textbf{1.7}} & 1.74 \\
    RE-RV-NN & 1.2   & \textit{\textbf{2.7}} & \textit{\textbf{2.3}} & \textit{\textbf{1.8}} & \textit{\textbf{2.1}} & \textit{\textbf{2.3}} & \textit{\textbf{2.6}} & 2.14 \\
    RE-RV-SkN & 0.7   & 1.6   & 1.5   & 1.3   & 1.3   & 1.5   & 1.3   & 1.31 \\
    RE-RV-tN & \cellcolor[rgb]{ .573,  .816,  .314}\textbf{1.0} & 1.5   & 1.5   & \textbf{0.9} & 1.3   & 1.3   & 1.3   & 1.26 \\
    RE-RR-NN & \textbf{1.1} & \textit{\textbf{2.7}} & \textit{\textbf{1.7}} & 1.4   & \textit{\textbf{1.9}} & \textit{\textbf{2.6}} & 2.2   & 1.94 \\
    RE-RR-SkN & 0.7   & 1.3   & \cellcolor[rgb]{ .573,  .816,  .314}\textbf{1.1} & \textbf{0.9} & \cellcolor[rgb]{ .573,  .816,  .314}\textbf{0.9} & 1.5   & \cellcolor[rgb]{ .573,  .816,  .314}\textbf{1.0} & 1.06 \\
    RE-RR-tN & 0.8   & 1.4   & 1.2   & 0.5   & 0.8   & 1.3   & \cellcolor[rgb]{ .573,  .816,  .314}\textbf{1.0} & 1.00 \\
    RE-RK-NN & 1.6   & \textit{\textbf{2.7}} & \textit{\textbf{1.9}} & 1.6   & \textit{\textbf{2.3}} & \textit{\textbf{1.7}} & \textit{\textbf{2.1}} & 1.99 \\
    RE-RK-SkN & \textbf{1.1} & 1.4   & \cellcolor[rgb]{ .573,  .816,  .314}\textbf{1.1} & \textbf{1.1} & 2.3   & 0.7   & 0.6   & 1.19 \\
    RE-RK-tN & 0.7   & 1.6   & 1.2   & \textbf{1.1} & 1.3   & 0.5   & 0.6   & 1.00 \\
    RE-RV-RR-NN & \textbf{1.1} & \textit{\textbf{2.8}} & \textit{\textbf{1.8}} & 1.5   & \textit{\textbf{1.9}} & \textit{\textbf{2.5}} & \textit{\textbf{2.2}} & 1.97 \\
    RE-RV-RR-SkN & 0.6   & 1.4   & 1.2   & \textbf{1.1} & 1.2   & 1.5   & \textbf{0.9} & 1.13 \\
    RE-RV-RR-tN & 0.7   & 1.6   & 1.4   & 0.5   & 1.2   & 1.2   & 0.6   & 1.03 \\
    RE-RV-RK-NN & 1.6   & \textit{\textbf{2.8}} & \textit{\textbf{2.1}} & \textit{\textbf{1.7}} & \textit{\textbf{2.2}} & \textit{\textbf{2.2}} & \textit{\textbf{2.4}} & 2.14 \\
    RE-RV-RK-SkN & 1.2   & 2.8   & 1.4   & 1.3   & 1.3   & 1.4   & 1.2   & 1.51 \\
    RE-RV-RK-tN & 1.2   & 1.5   & 1.4   & \cellcolor[rgb]{ .573,  .816,  .314}\textbf{1.0} & 1.3   & 1.2   & \cellcolor[rgb]{ .573,  .816,  .314}\textbf{1.0} & 1.23 \\
    RE-RR-RK-NN & 1.4   & \textit{\textbf{2.8}} & \textit{\textbf{1.7}} & 1.4   & \textit{\textbf{2.1}} & \textit{\textbf{2.5}} & \textit{\textbf{1.9}} & 1.97 \\
    RE-RR-RK-SkN & \textbf{0.9} & 1.4   & \cellcolor[rgb]{ .573,  .816,  .314}\textbf{1.1} & \cellcolor[rgb]{ .573,  .816,  .314}\textbf{1.0} & 1.2   & 1.4   & \textbf{1.1} & 1.16 \\
    RE-RR-RK-tN & \textbf{1.1} & 1.4   & \cellcolor[rgb]{ .573,  .816,  .314}\textbf{0.9} & 0.5   & \cellcolor[rgb]{ .573,  .816,  .314}\textbf{1.1} & 1.2   & 0.7   & 0.99 \\
    RE-RV-RR-RK-NN & 1.3   & \textit{\textbf{2.8}} & \textit{\textbf{1.7}} & 1.5   & \textit{\textbf{2.0}} & \textit{\textbf{2.4}} & \textit{\textbf{2.3}} & 2.00 \\
    RE-RV-RR-RK-SkN & \cellcolor[rgb]{ .573,  .816,  .314}\textbf{1.0} & \textit{\textbf{1.8}} & \cellcolor[rgb]{ .573,  .816,  .314}\textbf{1.1} & \cellcolor[rgb]{ .573,  .816,  .314}\textbf{1.0} & 1.2   & 1.3   & \cellcolor[rgb]{ .573,  .816,  .314}\textbf{1.0} & 1.20 \\
    RE-RV-RR-RK-tN & \cellcolor[rgb]{ .573,  .816,  .314}\textbf{1.0} & 1.4   & 1.4   & 0.5   & \cellcolor[rgb]{ .573,  .816,  .314}\textbf{1.1} & \cellcolor[rgb]{ .573,  .816,  .314}\textbf{1.0} & \textbf{0.9} & 1.04 \\
    \bottomrule
    \multicolumn{9}{p{45em}}{Note: For individual models, green highlight indicates the favoured model, bold indicates the 2nd ranked model, bold italics indicates that the violation rate is significantly different to 1\% by the UC test.}\\
    \end{tabular}%
}
\end{center}
  \label{tab:addlabel}%
\end{table}%

The result of three types of VaR backtests (the UC, CC, and DQ tests) conducted at 5\% significance level is presented in Table \ref{table:VaR_backtest}. The proposed RE-SkN and RE-tN type models have a lower number of rejections than the competing models, as highlighted in green. The standard GARCH-family models have relatively high rejection rates, as expected. The RE-NN type models tend to also have a relatively high number of rejections. Meanwhile, most of the RE-SkN and RE-tN type models are not rejected in the backtests for all indices. Specifically, the 2.5\% VaR forecasts generated by the RE-RK-SkN, RE-RV-RK-tN, and RE-RV-RR-RK-SkN models are not rejected for all three types of tests. For the 1\% VaR forecasts, the RE-RK-SkN and RE-RR-SkN models have no rejection for all three types of tests. This finding is consistent across all types of realized measures, either using a single or multiple realized measures.   

\begin{table}[H]
  \begin{center} 
  \captionsetup{font=normalsize, skip=0pt, justification= centering}
  \caption{VaR Backtest: UC, CC, and DQ tests}
  \label{table:VaR_backtest}
    \resizebox{0.6\textwidth}{!}{
    \begin{tabular}{lcccccc}
    \toprule
    \multicolumn{1}{c}{\multirow{2}[4]{*}{\textbf{Model}}} & \multicolumn{3}{c}{$\alpha$ = 2.5\%} & \multicolumn{3}{c}{$\alpha$  = 1\%} \\
\cmidrule{2-7}          & UC    & CC    & DQ4   & UC    & CC    & DQ4 \\
    \midrule
    GARCH-t & 7     & 4     & 6     & 4     & 4     & 5 \\
    EGARCH-t & 3     & 1     & 3     & 2     & 1     & 4 \\
    GJR-t & 4     & 3     & 2     & 2     & 1     & 3 \\
    GARCH-t-HS & \cellcolor[rgb]{ .573,  .816,  .314}\textbf{0} & 7     & 7     & \cellcolor[rgb]{ .573,  .816,  .314}\textbf{0} & 7     & 7 \\
    GARCH-t-EVT & 3     & 3     & 6     & 4     & 3     & 6 \\
    CARE-SAV & \cellcolor[rgb]{ .573,  .816,  .314}\textbf{0} & 1     & 3     & \cellcolor[rgb]{ .573,  .816,  .314}\textbf{0} & 2     & 2 \\
    RG-RV-tN & 5     & 4     & 5     & 6     & 6     & 6 \\
    RG-RR-tN & 3     & 3     & 3     & 3     & 3     & 5 \\
    RG-RK-tN & 5     & 4     & 5     & 4     & 5     & 5 \\
    RE-RV-NN & 4     & 2     & 4     & 6     & 5     & 5 \\
    RE-RV-SkN & \cellcolor[rgb]{ .573,  .816,  .314}\textbf{0} & \cellcolor[rgb]{ .573,  .816,  .314}\textbf{0} & 1     & \cellcolor[rgb]{ .573,  .816,  .314}\textbf{0} & \cellcolor[rgb]{ .573,  .816,  .314}\textbf{0} & 2 \\
    RE-RV-TN & \cellcolor[rgb]{ .573,  .816,  .314}\textbf{0} & \cellcolor[rgb]{ .573,  .816,  .314}\textbf{0} & 1     & \cellcolor[rgb]{ .573,  .816,  .314}\textbf{0} & \cellcolor[rgb]{ .573,  .816,  .314}\textbf{0} & 2 \\
    RE-RR-NN & 2     & 2     & 2     & 4     & 3     & 4 \\
    RE-RR-SkN & \cellcolor[rgb]{ .573,  .816,  .314}\textbf{0} & \cellcolor[rgb]{ .573,  .816,  .314}\textbf{0} & 2     & \cellcolor[rgb]{ .573,  .816,  .314}\textbf{0} & \cellcolor[rgb]{ .573,  .816,  .314}\textbf{0} & \cellcolor[rgb]{ .573,  .816,  .314}\textbf{0} \\
    RE-RR-TN & 1     & 1     & 1     & \cellcolor[rgb]{ .573,  .816,  .314}\textbf{0} & \cellcolor[rgb]{ .573,  .816,  .314}\textbf{0} & 1 \\
    RE-RK-NN & 3     & 2     & 3     & 5     & 4     & 5 \\
    RE-RK-SkN & \cellcolor[rgb]{ .573,  .816,  .314}\textbf{0} & \cellcolor[rgb]{ .573,  .816,  .314}\textbf{0} & \cellcolor[rgb]{ .573,  .816,  .314}\textbf{0} & \cellcolor[rgb]{ .573,  .816,  .314}\textbf{0} & \cellcolor[rgb]{ .573,  .816,  .314}\textbf{0} & \cellcolor[rgb]{ .573,  .816,  .314}\textbf{0} \\
    RE-RK-TN & 1     & 1     & \cellcolor[rgb]{ .573,  .816,  .314}\textbf{0} & \cellcolor[rgb]{ .573,  .816,  .314}\textbf{0} & \cellcolor[rgb]{ .573,  .816,  .314}\textbf{0} & 1 \\
    RE-RV-RR-NN & 3     & 2     & 2     & 5     & 5     & 5 \\
    RE-RV-RR-SkN & \cellcolor[rgb]{ .573,  .816,  .314}\textbf{0} & \cellcolor[rgb]{ .573,  .816,  .314}\textbf{0} & 1     & \cellcolor[rgb]{ .573,  .816,  .314}\textbf{0} & \cellcolor[rgb]{ .573,  .816,  .314}\textbf{0} & 2 \\
    RE-RV-RR-TN & 1     & \cellcolor[rgb]{ .573,  .816,  .314}\textbf{0} & 1     & \cellcolor[rgb]{ .573,  .816,  .314}\textbf{0} & \cellcolor[rgb]{ .573,  .816,  .314}\textbf{0} & 2 \\
    RE-RV-RK-NN & 4     & 3     & 4     & 6     & 5     & 6 \\
    RE-RV-RK-SkN & \cellcolor[rgb]{ .573,  .816,  .314}\textbf{0} & \cellcolor[rgb]{ .573,  .816,  .314}\textbf{0} & 1     & \cellcolor[rgb]{ .573,  .816,  .314}\textbf{0} & \cellcolor[rgb]{ .573,  .816,  .314}\textbf{0} & 2 \\
    RE-RV-RK-TN & \cellcolor[rgb]{ .573,  .816,  .314}\textbf{0} & \cellcolor[rgb]{ .573,  .816,  .314}\textbf{0} & \cellcolor[rgb]{ .573,  .816,  .314}\textbf{0} & \cellcolor[rgb]{ .573,  .816,  .314}\textbf{0} & \cellcolor[rgb]{ .573,  .816,  .314}\textbf{0} & 2 \\
    RE-RR-RK-NN & 3     & 2     & 3     & 5     & 4     & 5 \\
    RE-RR-RK-SkN & \cellcolor[rgb]{ .573,  .816,  .314}\textbf{0} & \cellcolor[rgb]{ .573,  .816,  .314}\textbf{0} & 2     & \cellcolor[rgb]{ .573,  .816,  .314}\textbf{0} & \cellcolor[rgb]{ .573,  .816,  .314}\textbf{0} & 1 \\
    RE-RR-RK-TN & 1     & \cellcolor[rgb]{ .573,  .816,  .314}\textbf{0} & 1     & \cellcolor[rgb]{ .573,  .816,  .314}\textbf{0} & \cellcolor[rgb]{ .573,  .816,  .314}\textbf{0} & 2 \\
    RE-RV-RR-RK-NN & 3     & 3     & 4     & 5     & 4     & 5 \\
    RE-RV-RR-RK-SkN & \cellcolor[rgb]{ .573,  .816,  .314}\textbf{0} & \cellcolor[rgb]{ .573,  .816,  .314}\textbf{0} & \cellcolor[rgb]{ .573,  .816,  .314}\textbf{0} & 1     & 1     & 2 \\
    RE-RV-RR-RK-TN & \cellcolor[rgb]{ .573,  .816,  .314}\textbf{0} & \cellcolor[rgb]{ .573,  .816,  .314}\textbf{0} & 1     & \cellcolor[rgb]{ .573,  .816,  .314}\textbf{0} & \cellcolor[rgb]{ .573,  .816,  .314}\textbf{0} & 1 \\
    \midrule
    \multicolumn{7}{p{25em}}{Note: For each test, green highlight indicates the model with the least  number of rejections} \\
    
    \end{tabular}%
}    
\end{center}
  \label{tab:addlabel}%
\end{table}%

Table \ref{table:ESR_backtest} shows the result of the bivariate ESR backtests \citep{BayerDimitriadis2018} in equation (\ref{bivariate_ESR}) across all market indices at the 5\% significance level. There is a small number of rejection for the null hypothesis of correct ES forecasts by almost all ESR backtests, even for the traditional GARCH-family models employing the Student-\textit{t} distribution, except for the GARCH-t-EVT model. This finding is consistent with the empirical studies conducted by \citet{BayerDimitriadis2018} that GARCH and GJR-GARCH models employing Student-\textit{t} and skewed Student-\textit{t} are not rejected by the ESR backtests. From Table \ref{table:ESR_backtest}, we can see the difference in performance across models at the 1\% forecasting level, where the proposed RE-tN and RE-SkN class models could generate more accurate ES forecasts compared to the competing models and the RE-NN class models. All proposed RE-tN and RE-SkN models have 0 ESR backtest rejection, except for the RE-RR-SkN and RE-RK-tN models. The best performing models with only 1 rejection for the 2.5\% ES forecasts and 0 rejection for the 1\% ES forecasts are the RE-RV-SkN, RE-RV-RK-SkN, RE-RR-RK-SkN, RE-RV-RR-RK-SkN, and RE-RV-RR-RK-tN models.

\begin{table}[H]
    \begin{center}
    \captionsetup{font=normalsize, skip=0pt, justification= centering}
    \caption{ES regression-based backtest (ESR)}
    \label{table:ESR_backtest}
  \resizebox{0.6\textwidth}{!}{
    \begin{tabular}{lcc}
    \toprule
    \multicolumn{1}{c}{\multirow{2}[4]{*}{\textbf{Model}}} & \multicolumn{2}{c}{\textbf{Total Rejections}} \\
\cmidrule{2-3}          & \textbf{$\alpha$ =2.5\%} & \textbf{$\alpha$ =1\%} \\
    \midrule
    GARCH-t & 1     & 1 \\
    EGARCH-t & 2     & 1 \\
    GJR-t & 1     & 1 \\
    GARCH-t-HS & 1     & 1 \\
    GARCH-t-EVT & 7     & 5 \\
    CARE-SAV & 4     & 3 \\
    RG-RV-tN & 3     & 1 \\
    RG-RR-tN & 1     & 1 \\
    RG-RK-tN & 2     & 1 \\
    RE-RV-NN & 3     & 1 \\
    RE-RV-SkN & 1     & 0 \\
    RE-RV-tN & 2     & 0 \\
    RE-RR-NN & 2     & 0 \\
    RE-RR-SkN & 1     & 1 \\
    RE-RR-tN & 2     & 0 \\
    RE-RK-NN & 2     & 2 \\
    RE-RK-SkN & 2     & 0 \\
    RE-RK-tN & 2     & 1 \\
    RE-RV-RR-NN & 2     & 0 \\
    RE-RV-RR-SkN & 2     & 0 \\
    RE-RV-RR-tN & 2     & 0 \\
    RE-RV-RK-NN & 3     & 0 \\
    RE-RV-RK-SkN & 1     & 0 \\
    RE-RV-RK-tN & 2     & 0 \\
    RE-RR-RK-NN & 2     & 0 \\
    RE-RR-RK-SkN & 1     & 0 \\
    RE-RR-RK-tN & 2     & 0 \\
    RE-RV-RR-RK-NN & 3     & 0 \\
    RE-RV-RR-RK-SkN & 1     & 0 \\
    RE-RV-RR-RK-tN & 1     & 0 \\
    \bottomrule
    \multicolumn{3}{p{25em}}{Note: Under the null hypothesis, the ES forecast are correctly specified. The tests are conducted at 5\% level of significance. Less rejection is favored.} \\    
    \end{tabular}%
}\\
\end{center}
\label{tab:addlabel}%
\end{table}%

\begin{figure}[h]
    \centering
    \captionsetup{font=normalsize, skip=0pt, justification= centering}
    \caption{6 quantiles VaR forecasts of the RE-RV-RK-SkN model for S\&P500}
    \label{fig:VaR_6quantiles_S&P}
      \resizebox{0.85\textwidth}{!}{
    \includegraphics[width=\linewidth]{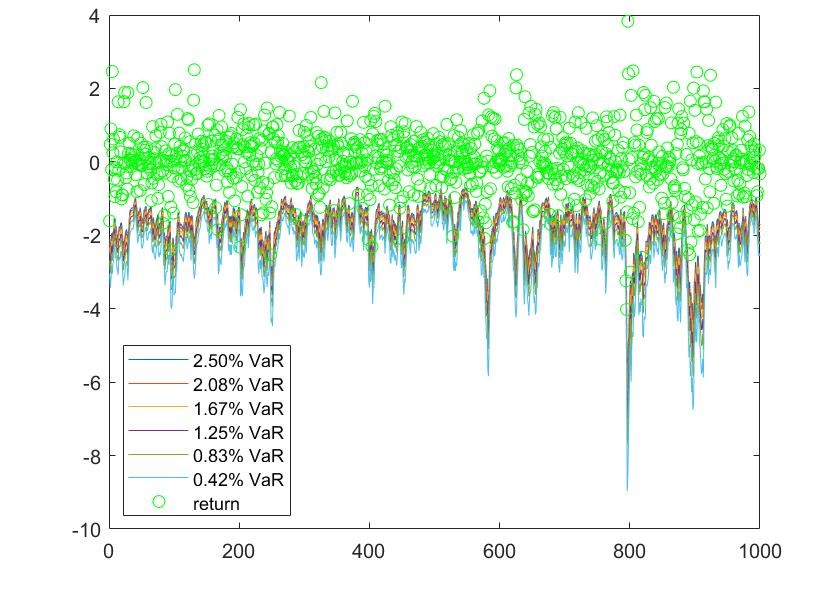}
}
\end{figure}

In Table \ref{table:ES_multiquantile_reg}, the result of the ES backtests based on the multi-quantile regression (ES-MQR) approach of \citet{CouperierLeymarie2019} in equation (\ref{ES_multiquantile_reg}) across models and all market indices is presented. The ES-MQR is conducted by using the number of quantile $p=6$ at coverage level $\tau=2.5\%$, as recommended by \citet{CouperierLeymarie2019}. As an example, the 6 quantiles VaR forecasts based on the RE-RV-RR-RK-SkN model for the S\&P500 used for the ES-MQR backtests are presented in Figure \ref{fig:VaR_6quantiles_S&P}.

\begin{table}[H]
\begin{center}
\captionsetup{font=normalsize, skip=0pt, justification= centering}
  \caption{ES backtests based on the multi-quantile regression ES-MQR) method}
  \label{table:ES_multiquantile_reg}
  \resizebox{0.75\textwidth}{!}{
    \begin{tabular}{lccccc}
    \toprule
    \multicolumn{1}{c}{\multirow{2}[4]{*}{\textbf{  Model  }}} & \multicolumn{4}{c}{\textbf{Hypothesis}} & \multicolumn{1}{c}{\multirow{2}[4]{*}{\textbf{Total Rejections}}} \\
\cmidrule{2-5}          & \textbf{  $J_1$  } & \textbf{  $J_2$  } & \textbf {  $I$  } & \textbf{  $S$  } &  \\
    \midrule
    GARCH-t & 4     & 4     & 4     & 4     & 16 \\
    EGARCH-t & 3     & 5     & 3     & 4     & 15 \\
    GJR-t & 4     & 3     & 4     & 5     & 16 \\
    GARCH-t-HS & 4     & 6     & 4     & 4     & \textit{18} \\
    GARCH-t-EVT & 7     & 7     & 7     & 7     & \textit{\textbf{28}} \\
    CARE-SAV & 2     & 7     & \cellcolor[rgb]{ .573,  .816,  .314}\textbf{0} & 4     & 13 \\
    RG-RV-tN & 3     & 3     & 2     & 3     & 11 \\
    RG-RR-tN & 2     & 3     & 1     & 3     & 9 \\
    RG-RK-tN & 1     & 4     & 2     & 1     & 8 \\
    RE-RV-NN & 1     & 2     & 1     & 2     & 6 \\
    RE-RV-SkN & 2     & 5     & \cellcolor[rgb]{ .573,  .816,  .314}\textbf{0} & 4     & 11 \\
    RE-RV-tN & 2     & 6     & 2     & 4     & 14 \\
    RE-RR-NN & 1     & 3     & 1     & 1     & 6 \\
    RE-RR-SkN & 2     & 4     & 1     & 3     & 10 \\
    RE-RR-tN & 3     & 6     & \cellcolor[rgb]{ .573,  .816,  .314}\textbf{0} & 3     & 12 \\
    RE-RK-NN & 1     & \cellcolor[rgb]{ .573,  .816,  .314}\textbf{1} & 1     & 1     & \textbf{4} \\
    RE-RK-SkN & 1     & 6     & 1     & 3     & 11 \\
    RE-RK-tN & 2     & 7     & 1     & 3     & 13 \\
    RE-RV-RR-NN & 1     & 2     & 1     & 1     & 5 \\
    RE-RV-RR-SkN & 1     & 4     & 1     & 3     & 9 \\
    RE-RV-RR-tN & 2     & 6     & 1     & 4     & 13 \\
    RE-RV-RK-NN & \cellcolor[rgb]{ .573,  .816,  .314}\textbf{0} & \cellcolor[rgb]{ .573,  .816,  .314}1 & \cellcolor[rgb]{ .573,  .816,  .314}\textbf{0} & 1     & \cellcolor[rgb]{ .573,  .816,  .314}\textbf{2} \\
    RE-RV-RK-SkN & \cellcolor[rgb]{ .573,  .816,  .314}\textbf{0} & 3     & \cellcolor[rgb]{ .573,  .816,  .314}\textbf{0} & 2     & 5 \\
    RE-RV-RK-tN & 2     & 5     & 1     & 2     & 10 \\
    RE-RR-RK-NN & \cellcolor[rgb]{ .573,  .816,  .314}\textbf{0} & 2     & \cellcolor[rgb]{ .573,  .816,  .314}\textbf{0} & \cellcolor[rgb]{ .573,  .816,  .314}\textbf{0} & \cellcolor[rgb]{ .573,  .816,  .314}\textbf{2} \\
    RE-RR-RK-SkN & \cellcolor[rgb]{ .573,  .816,  .314}\textbf{0} & 4     & \cellcolor[rgb]{ .573,  .816,  .314}\textbf{0} & 1     & 5 \\
    RE-RR-RK-tN & \cellcolor[rgb]{ .573,  .816,  .314}\textbf{0} & 5     & \cellcolor[rgb]{ .573,  .816,  .314}\textbf{0} & 2     & 7 \\
    RE-RV-RR-RK-NN & \cellcolor[rgb]{ .573,  .816,  .314}\textbf{0} & \cellcolor[rgb]{ .573,  .816,  .314}\textbf{1} & \cellcolor[rgb]{ .573,  .816,  .314}\textbf{0} & 1     & \cellcolor[rgb]{ .573,  .816,  .314}\textbf{2} \\
    RE-RV-RR-RK-SkN & \cellcolor[rgb]{ .573,  .816,  .314}\textbf{0} & 4     & \cellcolor[rgb]{ .573,  .816,  .314}\textbf{0} & 3     & 7 \\
    RE-RV-RR-RK-tN & 1     & 6     & \cellcolor[rgb]{ .573,  .816,  .314}\textbf{0} & 1     & 8 \\
    \bottomrule
    \multicolumn{6}{p{30em}}{Note:  Green highlight indicates the favoured model, bold indicates the 2nd ranked model, bold italics indicates the least favoured model, italics indicates the 2nd lowest ranked model in each column.}     
    \end{tabular}%
}
\end{center}
  \label{tab:addlabel}%
\end{table}%

All four hypothesis testings of the ES-MQR backtests are conducted at the 5\% significance level based on bootstrap critical values. With respect to the total rejections of 4 hypotheses of valid ES forecasts for 7 market indices, the ES-MQR backtests of the realized EGARCH models estimated using the adaptive MCMC approach (the RE-NN,RE-tN, RE-SkN class models) are less likely to be rejected than the standard GARCH-type models using the Student-\textit{t} distribution. However, some of those proposed models only perform slightly better than the CARE-SAV and RG-tN models, or even slightly worse than the RG-tN models. For instance, the RE-RV-tN model has a higher total number of rejections compared to the CARE-SAV and RG-tN models. 

In terms of the individual hypotheses testings, the first joint backtests that sums the the intercept and the slope coefficients together ($H_{0,J_1}$) and the intercept backtests ($H_{0,I}$) for the realized EGARCH models are less likely to reject the validity of ES forecasts. The non-rejection of $H_{0,I}$ implies that the forecasting errors are not constant across time. The other joint backtests that sum the intercept and the slope coefficients separately ($H_{0,J_2}$) tend to reject the validity of ES forecasts in most models. This is due to the rejection of the slope backtests ($H_{0,S}$), which indicates that the errors are time-varying since they fluctuate with respect to the VaR forecasts.

While the RE-SkN type models are relatively preferable than the RE-tN type models, the ES-MQR backtests surprisingly are less likely to reject the tail risk validity produced by the RE-NN class models compared to the RE-SkN and RE-tN class models. This is rather contradictory to the result of the ESR backtests which provide more supports for the RE-SkN and RE-tN type models. However, the rejection of the ES-MQR backtests for the RE-SkN and RE-tN models are mainly due to the rejection of $H_{0,J_2}$ and $H_{0,S}$ hypotheses. For the other two ES backtests hypothesis tests ($H_{0,J_1}$ and $H_{0,I}$), the RE-SkN and RE-tN type models perform as good as the RE-NN type models. Lastly, the realized EGARCH models with multiple realized measures have a lower ES-MQR backtest rejections than those with single realized measure. 

\subsection{Quantile loss functions}

In Table \ref{table:quantile_loss_25pct} and \ref{table:quantile_loss_1pct}, we present the quantile loss function values at 2.5\% and 1\% forecasting levels, respectively. The proposed RE-SkN and RE-tN type models in general generate smaller quantile losses and rank better compared to other models and the RE-NN class models. 

% Cek violation rate

For the 2.5\% VaR forecast, the best model generating the lowest average quantile loss value is the RE-RV-RK-SkN model (63.215), followed by the RE-RV-RR-RK-SkN model (63.220). In terms of average rank, the RE-RV-RK-SkN model also ranks the best, followed by the RE-RV-RK-tN model. On the contrary, the GARCH-t-EVT and GARCH-t models produce the highest average quantile loss values, which is consistent with the VRate of these models in Table \ref{table:VRate25pct}. On the one hand, the GARCH-t-EVT model tends to underestimate risks with an average VRate of 1.43\%, which is much lower than the nominal VRate of 2.5\%. On the other hand, the GARCH-t model overestimate risks, producing VRate much higher than 2.5\%, i.e. 3.76\%.    

For the 1\% forecast, the model that has the lowest average quantile loss value (29.541) and ranks the best (5.4 on average) is the RE-RV-RR-RK-SkN model. As has been discussed, the VRate of this model is also the best for 4 out of 7 market indices in this study, although its average VRate across 7 market indices is not the lowest. The second best model is the RE-RV-RK-SkN model with the average quantile loss value of 29.643, while the RE-RV-RR-SkN model has the second best average rank (7.3). It is evident for both 2.5\% and 1\% forecasting levels that the proposed RE-SkN type models are more preferred in VaR forecasting.

\begin{table}[H]
  \begin{center}
  \captionsetup{font=normalsize, skip=0pt, justification= centering}
  \caption{Quantile loss function values; $\alpha=2.5\%$}
  \label{table:quantile_loss_25pct}
  \resizebox{1.15\textwidth}{!}{
    \begin{tabular}{lccccccccc}
    \toprule
    \multicolumn{1}{c}{\textbf{Model}} & \textbf{ASX200} & \textbf{DAX} & \textbf{FTSE} & \textbf{HK} & \textbf{NASDAQ} & \textbf{SMI} & \textbf{S\&P500} & \textbf{Mean Loss} & \textbf{Mean Rank} \\
    \midrule
    GARCH-t & \textit{57.801} & 80.213 & \textit{62.499} & 76.265 & 66.537 & 69.750 & 56.700 & \textit{67.103} & \textit{27.0} \\
    EGARCH-t & 54.925 & 78.204 & 59.180 & 73.726 & 63.206 & 68.774 & 53.840 & 64.550 & 19.3 \\
    GJR-t & 56.707 & 79.343 & 59.779 & 73.114 & 63.810 & 68.708 & 54.000 & 65.066 & 22.3 \\
    GARCH-t-HS & 57.756 & 80.133 & 62.454 & 75.875 & 66.424 & 69.844 & 56.700 & 67.027 & 26.4 \\
    GARCH-t-EVT & \textit{\textbf{90.850}} & \textit{\textbf{108.287}} & \textit{\textbf{86.737}} & \textit{\textbf{91.598}} & \textit{\textbf{72.950}} & \textit{\textbf{90.156}} & \textit{\textbf{67.292}} & \textit{\textbf{86.839}} & \textit{\textbf{30.0}} \\
    CARE-SAV & 55.334 & 78.659 & 59.965 & \textit{76.349} & 65.408 & 72.118 & \textit{57.845} & 66.526 & 25.0 \\
    RG-RV-tN & 54.166 & \textit{82.657} & 60.951 & 72.225 & 64.966 & 69.571 & 55.521 & 65.723 & 22.7 \\
    RG-RR-tN & 54.156 & 81.807 & 59.683 & 71.781 & 64.224 & 70.438 & 54.264 & 65.193 & 20.9 \\
    RG-RK-tN & 55.098 & 81.962 & 61.489 & 73.008 & \textit{66.660} & \textit{75.288} & 55.097 & 66.943 & 26.0 \\
    RE-RV-NN & 53.699 & 79.982 & 59.452 & 71.985 & 62.160 & 67.764 & 53.254 & 64.042 & 16.9 \\
    RE-RV-SkN & 53.332 & 78.604 & 58.022 & 71.824 & 62.090 & 67.034 & 52.091 & 63.285 & 8.7 \\
    RE-RV-TN & 53.519 & 78.514 & 57.899 & 72.241 & 61.884 & 67.207 & 52.038 & 63.329 & 9.0 \\
    RE-RR-NN & 53.613 & 79.792 & 58.678 & \cellcolor[rgb]{ .573,  .816,  .314}\textbf{71.569} & 62.005 & 68.185 & 52.879 & 63.817 & 14.3 \\
    RE-RR-SkN & 53.739 & 78.936 & \textbf{57.547} & 71.975 & 62.152 & 67.599 & 52.127 & 63.439 & 11.0 \\
    RE-RR-TN & 53.390 & 78.577 & 57.752 & 74.610 & 62.013 & \cellcolor[rgb]{ .573,  .816,  .314}\textbf{66.993} & 52.320 & 63.665 & 11.1 \\
    RE-RK-NN & 54.447 & 79.329 & 59.070 & 72.537 & 62.742 & 68.530 & 52.383 & 64.148 & 18.4 \\
    RE-RK-SkN & 54.373 & \textbf{78.117} & 58.089 & 72.280 & 62.317 & 70.062 & 52.056 & 63.899 & 14.7 \\
    RE-RK-TN & 54.746 & \cellcolor[rgb]{ .573,  .816,  .314}\textbf{77.736} & 58.008 & 73.024 & 62.089 & 73.830 & 52.649 & 64.583 & 16.1 \\
    RE-RV-RR-NN & 53.446 & 79.815 & 58.846 & 72.016 & 61.962 & 68.169 & 52.821 & 63.868 & 14.6 \\
    RE-RV-RR-SkN & 54.085 & 79.033 & 57.728 & 72.343 & 61.917 & 67.620 & 51.832 & 63.509 & 11.6 \\
    RE-RV-RR-TN & 53.558 & 78.577 & 57.610 & 73.943 & 61.867 & 67.276 & \textbf{51.379} & 63.459 & 9.4 \\
    RE-RV-RK-NN & 53.509 & 79.804 & 59.171 & 72.176 & 62.366 & 67.595 & 52.870 & 63.927 & 15.9 \\
    RE-RV-RK-SkN & 53.241 & 78.329 & 57.592 & 72.277 & 61.785 & 67.165 & 52.118 & \cellcolor[rgb]{ .573,  .816,  .314}\textbf{63.215} & \cellcolor[rgb]{ .573,  .816,  .314}\textbf{6.4} \\
    RE-RV-RK-TN & 53.345 & 78.397 & 57.780 & 72.922 & 61.642 & \textbf{67.030} & \cellcolor[rgb]{ .573,  .816,  .314}\textbf{51.490} & 63.229 & \textbf{6.6} \\
    RE-RR-RK-NN & 53.396 & 80.086 & 58.403 & 72.002 & 61.636 & 67.816 & 52.544 & 63.698 & 12.6 \\
    RE-RR-RK-SkN & \textbf{53.161} & 78.558 & 57.783 & \textbf{71.747} & 62.122 & 67.688 & 52.353 & 63.345 & 9.0 \\
    RE-RR-RK-TN & 53.355 & 78.536 & \cellcolor[rgb]{ .573,  .816,  .314}\textbf{57.328} & 74.310 & \textbf{61.632} & 67.338 & 51.556 & 63.437 & 7.7 \\
    RE-RV-RR-RK-NN & 53.165 & 80.139 & 58.833 & 72.020 & 62.253 & 67.841 & 52.839 & 63.870 & 15.1 \\
    RE-RV-RR-RK-SkN & 53.251 & 78.859 & 57.684 & 72.228 & \cellcolor[rgb]{ .573,  .816,  .314}\textbf{61.261} & 67.420 & 51.838 & \textbf{63.220} & 7.3 \\
    RE-RV-RR-RK-TN & \cellcolor[rgb]{ .573,  .816,  .314}\textbf{53.145} & 78.552 & 57.801 & 74.028 & 61.701 & 67.201 & 52.083 & 63.502 & 8.7 \\
    \bottomrule
    \multicolumn{10}{p{53em}}{Note: For individual models, green highlight indicates the favoured model, bold indicates the 2nd ranked model, bold italics indicates the least favoured model, italics indicates the 2nd lowest ranked model, in each column.} \\
    \end{tabular}%
}
\end{center}
  \label{tab:addlabel}%
\end{table}%

\begin{table}[H]
    \begin{center}
  \captionsetup{font=normalsize, skip=0pt, justification= centering}
  \caption{Quantile loss function values; $\alpha=1\%$}
  \label{table:quantile_loss_1pct}
    \resizebox{1.15\textwidth}{!}{%
    \begin{tabular}{lccccccccc}
    \toprule
    \multicolumn{1}{c}{\textbf{Model}} & \textbf{ASX200} & \textbf{DAX} & \textbf{FTSE} & \textbf{HK} & \textbf{NASDAQ} & \textbf{SMI} & \textbf{S\&P500} & \textbf{Mean Loss} & \textbf{Mean Rank} \\
    \midrule
    GARCH-t & \textit{26.731} & 36.925 & 29.449 & 36.873 & 29.736 & 35.759 & 26.410 & 31.698 & 23.7 \\
    EGARCH-t & 25.572 & 38.023 & 27.965 & 35.068 & 28.153 & 35.976 & 24.696 & 30.779 & 17.7 \\
    GJR-t & 25.772 & 38.542 & 27.581 & 34.493 & 28.474 & 36.222 & 24.720 & 30.829 & 18.7 \\
    GARCH-t-HS & \textit{26.731} & 36.526 & \textit{29.561} & 36.510 & 29.717 & 35.802 & 26.410 & 31.608 & 23.6 \\
    GARCH-t-EVT & \textit{\textbf{37.392}} & \textit{\textbf{49.205}} & \textit{\textbf{40.302}} & \textit{\textbf{39.483}} & \textit{\textbf{33.299}} & \textit{\textbf{47.263}} & \textit{\textbf{30.225}} & \textit{\textbf{39.596}} & \textit{\textbf{30.0}} \\
    CARE-SAV & 25.362 & \cellcolor[rgb]{ .573,  .816,  .314}\textbf{35.768} & 29.088 & \textit{37.863} & 30.125 & 37.272 & \textit{27.199} & 31.811 & 22.3 \\
    RG-RV-tN & 25.007 & \textit{39.541} & 29.222 & 34.610 & 30.344 & 37.100 & 26.679 & 31.786 & 23.4 \\
    RG-RR-tN & 25.086 & 38.481 & 28.128 & \textbf{33.116} & 29.329 & 37.810 & 25.252 & 31.029 & 19.0 \\
    RG-RK-tN & 26.214 & 38.794 & 29.262 & 34.777 & \textit{31.920} & \textit{40.033} & 26.400 & \textit{32.486} & \textit{26.4} \\
    RE-RV-NN & 24.943 & 38.711 & 28.378 & 35.116 & 29.267 & 36.626 & 25.437 & 31.211 & 21.6 \\
    RE-RV-SkN & 24.876 & 35.926 & 26.976 & 33.231 & 28.452 & 34.398 & 23.735 & 29.656 & 8.0 \\
    RE-RV-TN & \textbf{24.828} & 36.122 & 26.941 & 33.584 & 28.308 & 34.580 & 23.622 & 29.712 & 8.6 \\
    RE-RR-NN & 24.943 & 38.154 & 27.932 & 33.827 & 28.659 & 37.001 & 24.584 & 30.728 & 16.6 \\
    RE-RR-SkN & 25.060 & 35.888 & \textbf{26.705} & 33.276 & 28.427 & 34.554 & 23.855 & 29.681 & 8.3 \\
    RE-RR-TN & 24.923 & \textbf{35.808} & 26.832 & 35.024 & 28.385 & \cellcolor[rgb]{ .573,  .816,  .314}\textbf{34.128} & 23.911 & 29.858 & 9.3 \\
    RE-RK-NN & 26.082 & 38.280 & 28.103 & 35.040 & 29.244 & 36.500 & 24.074 & 31.046 & 20.9 \\
    RE-RK-SkN & 25.505 & 36.129 & 27.066 & 34.351 & 28.172 & 36.665 & 23.611 & 30.214 & 13.9 \\
    RE-RK-TN & 25.855 & 36.242 & 27.082 & 34.080 & 28.081 & 38.816 & 23.922 & 30.582 & 15.9 \\
    RE-RV-RR-NN & 24.834 & 38.541 & 27.985 & 34.482 & 28.514 & 36.915 & 24.493 & 30.824 & 17.0 \\
    RE-RV-RR-SkN & 25.267 & 36.011 & 26.812 & 33.555 & 28.036 & 34.817 & \cellcolor[rgb]{ .573,  .816,  .314}\textbf{23.205} & 29.672 & \textbf{7.3} \\
    RE-RV-RR-TN & 24.917 & 36.237 & 26.842 & 34.658 & 28.164 & 34.382 & 23.396 & 29.799 & 8.4 \\
    RE-RV-RK-NN & 25.873 & 38.653 & 28.166 & 34.859 & 29.185 & 36.423 & 24.772 & 31.133 & 22.1 \\
    RE-RV-RK-SkN & 25.104 & 36.011 & 26.711 & 33.571 & 28.292 & 34.405 & 23.404 & \textbf{29.643} & 7.7 \\
    RE-RV-RK-TN & 25.314 & 35.883 & 26.922 & 33.719 & 28.203 & 34.449 & 23.397 & 29.698 & 8.4 \\
    RE-RR-RK-NN & 25.559 & 38.588 & 27.622 & 34.300 & 28.382 & 36.587 & 24.215 & 30.751 & 17.6 \\
    RE-RR-RK-SkN & 25.012 & 36.003 & 26.934 & \cellcolor[rgb]{ .573,  .816,  .314}\textbf{33.012} & 28.330 & 34.793 & 23.701 & 29.683 & 9.0 \\
    RE-RR-RK-TN & 25.270 & 36.119 & \cellcolor[rgb]{ .573,  .816,  .314}\textbf{26.494} & 34.614 & 28.137 & \textbf{34.343} & 23.563 & 29.791 & 8.7 \\
    RE-RV-RR-RK-NN & \cellcolor[rgb]{ .573,  .816,  .314}\textbf{24.821} & 38.954 & 27.975 & 34.606 & 28.704 & 36.549 & 24.487 & 30.871 & 17.3 \\
    RE-RV-RR-RK-SkN & 25.025 & 35.869 & 26.817 & 33.443 & \cellcolor[rgb]{ .573,  .816,  .314}\textbf{27.534} & 34.725 & \textbf{23.372} & \cellcolor[rgb]{ .573,  .816,  .314}\textbf{29.541} & \cellcolor[rgb]{ .573,  .816,  .314}\textbf{5.4} \\
    RE-RV-RR-RK-TN & 24.977 & 35.986 & 26.896 & 33.935 & \textbf{28.010} & 34.664 & 23.630 & 29.728 & 7.9 \\
    \bottomrule
    \multicolumn{10}{p{53em}}{Note: For individual models, green highlight indicates the favoured model, bold indicates the 2nd ranked model, bold italics indicates the least favoured model, italics indicates the 2nd lowest ranked model, in each column.} \\    
    \end{tabular}%
}
\end{center}
  \label{tab:addlabel}%
\end{table}%

\subsection{Joint loss function of VaR and ES}
\label{section:joint_loss}

The values of the special case of the VaR and ES joint loss function of \citet{FisslerZiegel2016} based on equation (\ref{VaRES_jointloss_specific}) are presented in Table \ref{table:VaRES_joint_loss25pct} for 2.5\% forecasting level and Table \ref{table:VaRES_joint_loss1pct} for 1\% forecasting level. The proposed RE-SkN type models show a more desirable property compared to other models, generating lower VaR and ES joint loss values for 5 out of 7 market indices for either 2.5\% or 1\% forecasting level. The RE-RR-SkN model consistently produces the lowest average of the VaR and ES joint losses for both forecasting levels, followed by the RE-RR-RK-SkN model for the 2.5\% forecasting level and the RE-RV-RR-SkN model for the 1\% forecasting level. On average, the RE-RR-SkN  model also ranks first for the 1\% forecasting level and ranks second for the 2.5\% forecasting level (the RE-RR-RK-SkN model ranks first for the 2.5\% forecasting level). As regards the type of realized measures, the proposed RE-SkN type models employing RRSS, both individually or jointly with RVSS and/or RK, are more likely to have lower VaR and ES joint losses. We can conclude that the choice of the RRSS variable and the skewed Student-\textit{t} distribution as the return equation errors improves the tail risk forecasting efficiency.

\begin{table}[H]
  \begin{center}
  \captionsetup{font=normalsize, skip=0pt, justification= centering}
  \caption{VaR and ES joint loss function values based on equation (\ref{VaRES_jointloss_specific}); $\alpha=2.5\%$}
  \label{table:VaRES_joint_loss25pct}
  \resizebox{1.05\textwidth}{!}{
    \begin{tabular}{lccccccccc}
    \toprule
    \multicolumn{1}{c}{\textbf{Model}} & \textbf{ASX200} & \textbf{DAX} & \textbf{FTSE} & \textbf{HK} & \textbf{NASDAQ} & \textbf{SMI} & \textbf{S\&P500} & {\textbf{Mean Loss}} & {\textbf{Mean Rank}} \\
    \midrule
    GARCH-t & \textit{986.5} & 1068.8 & \textit{1004.8} & 1049.8 & 1015.6 & 1019.9 & 970.8 & 1016.6 & \textit{25.1} \\
    EGARCH-t & 970.7 & 1073.3 & 995.5 & 1042.6 & 1002.8 & 1036.1 & 948.2 & 1009.9 & 24.0 \\
    GJR-t & 980.2 & 1076.4 & 986.3 & 1037.2 & 1003.6 & 1030.2 & 951.1 & 1009.3 & 23.7 \\
    GARCH-t-HS & 965.2 & 1058.7 & 988.8 & 1045.8 & 1007.1 & 1012.1 & 962.9 & 1005.8 & 20.4 \\
    GARCH-t-EVT & \textit{\textbf{1085.0}} & \textit{\textbf{1128.6}} & \textit{\textbf{1092.2}} & \textit{\textbf{1091.6}} & \textit{\textbf{1035.5}} & \textit{\textbf{1157.3}} & \textit{\textbf{1018.4}} & \textit{\textbf{1086.9}} & \textit{\textbf{30.0}} \\
    CARE-SAV & 974.2 & \cellcolor[rgb]{ .573,  .816,  .314}\textbf{1057.6} & 990.9 & \textit{1054.7} & 1018.0 & 1037.5 & \textit{983.7} & \textit{1016.7} & 23.9 \\
    RG-RV-tN & 941.3 & \textit{1090.1} & 1001.7 & 1037.5 & 1009.4 & 1026.3 & 966.1 & 1010.3 & 24.0 \\
    RG-RR-tN & 941.2 & 1077.6 & 984.5 & 1028.7 & 996.4 & \textit{1041.1} & 945.4 & 1002.1 & 18.4 \\
    RG-RK-tN & 957.6 & 1083.1 & 1002.2 & 1031.8 & \textit{1026.7} & 1033.2 & 954.2 & 1012.7 & 24.9 \\
    RE-RV-NN & 943.7 & 1078.1 & 986.2 & 1035.4 & 998.4 & 1019.9 & 949.7 & 1001.6 & 20.7 \\
    RE-RV-SkN & 939.1 & 1060.4 & 969.4 & 1031.7 & 989.6 & \cellcolor[rgb]{ .573,  .816,  .314}\textbf{997.6} & 928.5 & 988.0 & 8.1 \\
    RE-RV-tN & 941.1 & 1066.6 & 972.7 & 1033.5 & 989.1 & 1002.4 & 928.7 & 990.6 & 12.6 \\
    RE-RR-NN & 941.7 & 1072.3 & 978.0 & \cellcolor[rgb]{ .573,  .816,  .314}\textbf{1028.5} & 991.4 & 1027.5 & 926.6 & 995.1 & 13.9 \\
    RE-RR-SkN & 940.6 & 1058.9 & \cellcolor[rgb]{ .573,  .816,  .314}\textbf{963.9} & 1030.5 & 985.8 & 1001.9 & 926.5 & \cellcolor[rgb]{ .573,  .816,  .314}\textbf{986.9} & \textbf{5.6} \\
    RE-RR-tN & 938.7 & 1061.1 & 968.6 & 1037.7 & 985.9 & 1003.4 & 927.9 & 989.0 & 9.6 \\
    RE-RK-NN & 957.2 & 1077.0 & 988.2 & 1033.1 & 1004.0 & 1016.4 & 931.2 & 1001.0 & 19.9 \\
    RE-RK-SkN & 951.6 & 1061.9 & 974.0 & 1029.8 & 992.5 & 1012.9 & \textbf{923.2} & 992.3 & 11.3 \\
    RE-RK-tN & 952.2 & 1063.5 & 976.1 & 1031.8 & 994.3 & 1028.4 & 926.6 & 996.1 & 16.0 \\
    RE-RV-RR-NN & 940.0 & 1076.0 & 979.0 & 1031.2 & 991.4 & 1026.5 & 937.7 & 997.4 & 15.7 \\
    RE-RV-RR-SkN & 943.4 & 1062.8 & 967.0 & 1031.1 & 986.5 & 1002.2 & 924.0 & 988.1 & 8.4 \\
    RE-RV-RR-tN & 940.8 & 1065.2 & 969.0 & 1036.5 & 987.1 & 1004.3 & 924.1 & 989.6 & 10.6 \\
    RE-RV-RK-NN & 947.4 & 1078.0 & 983.9 & 1034.0 & 998.9 & 1018.2 & 940.9 & 1000.2 & 19.6 \\
    RE-RV-RK-SkN & 940.2 & 1059.0 & 966.8 & 1029.6 & 990.5 & \textbf{998.2} & 925.9 & 987.2 & 6.1 \\
    RE-RV-RK-tN & 942.8 & 1065.4 & 972.2 & 1034.5 & 991.0 & 1000.6 & 925.4 & 990.3 & 11.7 \\
    RE-RR-RK-NN & 943.3 & 1076.7 & 974.5 & 1031.0 & 988.6 & 1022.0 & 933.8 & 995.7 & 15.1 \\
    RE-RR-RK-SkN & \cellcolor[rgb]{ .573,  .816,  .314}\textbf{936.2} & \textbf{1057.6} & 966.9 & \textbf{1028.6} & 988.7 & 1003.9 & 927.3 & \textbf{987.0} & 5.7 \\
    RE-RR-RK-tN & 940.5 & 1065.4 & \textbf{964.5} & 1037.0 & \textbf{984.8} & 1006.1 & \cellcolor[rgb]{ .573,  .816,  .314}\textbf{921.3} & 988.5 & 8.9 \\
    RE-RV-RR-RK-NN & 939.2 & 1079.4 & 980.1 & 1029.9 & 994.1 & 1022.8 & 941.8 & 998.2 & 16.1 \\
    RE-RV-RR-RK-SkN & 939.5 & 1064.7 & 966.5 & 1030.8 & \cellcolor[rgb]{ .573,  .816,  .314}\textbf{983.6} & 1001.3 & 924.0 & 987.2 & \cellcolor[rgb]{ .573,  .816,  .314}\textbf{5.1} \\
    RE-RV-RR-RK-tN & \textbf{937.4} & 1065.3 & 969.7 & 1035.5 & 986.2 & 1004.8 & 925.1 & 989.1 & 9.6 \\
    \bottomrule
    \multicolumn{10}{p{55em}}{Note: For individual models, green highlight indicates the favoured model, bold indicates the 2nd ranked model, bold italics indicates the least favoured model, italics indicates the 2nd lowest ranked model, in each column.} \\     
    \end{tabular}%
}
\end{center}
  \label{tab:addlabel}%
\end{table}%

    % Table generated by Excel2LaTeX from sheet 'Sheet4'
\begin{table}[H]
  \begin{center}
  \captionsetup{font=normalsize, skip=0pt, justification= centering}
  \caption{VaR and ES joint loss function values based on equation (\ref{VaRES_jointloss_specific}); $\alpha=1\%$}
  \label{table:VaRES_joint_loss1pct}
  \resizebox{1.1\textwidth}{!}{
    \begin{tabular}{lccccccccc}
    \toprule
    \multicolumn{1}{c}{\textbf{Model}} & \textbf{ASX200} & \textbf{DAX} & \textbf{FTSE} & \textbf{HK} & \textbf{NASDAQ} & \textbf{SMI} & \textbf{S\&P500} & \textbf{Mean Loss} & \textbf{Mean Rank} \\
    \midrule
    GARCH-t & 981.2 & 1042.2 & 982.0 & 1015.4 & 983.0 & 999.3 & 957.1 & 994.3 & 22.29 \\
    EGARCH-t & \textit{987.3} & 1064.7 & \textit{1006.9} & 1015.8 & 971.9 & 1043.6 & 938.6 & \textit{1004.1} & 24.57 \\
    GJR-t & 980.9 & 1059.5 & 974.5 & 1007.7 & 975.6 & 1034.1 & 938.2 & 995.8 & 20.86 \\
    GARCH-t-HS & 956.3 & 1032.8 & 977.1 & 1010.6 & 973.9 & 995.6 & 955.7 & 986.0 & 18.43 \\
    GARCH-t-EVT & \textit{\textbf{1033.8}} & \textit{\textbf{1120.7}} & \textit{\textbf{1116.0}} & \textit{\textbf{1040.2}} & \textit{\textbf{1050.1}} & \textit{\textbf{1247.3}} & \textit{\textbf{998.2}} & \textit{\textbf{1086.6}} & \textit{\textbf{30.00}} \\
    CARE-SAV & 963.5 & 1026.1 & 982.0 & \textit{1026.1} & 993.9 & 1012.9 & \textit{975.8} & 997.2 & 21.57 \\
    RG-RV-tN & \cellcolor[rgb]{ .573,  .816,  .314}\textbf{930.0} & 1062.5 & 996.2 & 1016.1 & 991.2 & 1032.5 & 961.4 & 998.6 & 21.57 \\
    RG-RR-tN & \textbf{931.6} & 1043.9 & 969.5 & 1001.0 & 970.5 & 1049.2 & 930.4 & 985.1 & 13.86 \\
    RG-RK-tN & 951.9 & 1052.4 & 992.6 & 1003.3 & \textit{1010.5} & 1014.7 & 944.9 & 995.7 & 21.29 \\
    RE-RV-NN & 944.3 & 1066.1 & 983.8 & 1021.3 & 997.2 & 1041.8 & 964.7 & 1002.8 & \textit{25.14} \\
    RE-RV-SkN & 940.4 & 1026.9 & 955.3 & 1005.3 & 969.3 & 985.5 & 916.8 & 971.4 & 9.86 \\
    RE-RV-tN & 941.5 & 1036.5 & 959.4 & 1004.8 & 967.5 & 993.9 & 911.8 & 973.6 & 12.14 \\
    RE-RR-NN & 940.7 & 1052.3 & 973.7 & 1006.3 & 979.4 & \textit{1050.5} & 912.1 & 987.9 & 18.00 \\
    RE-RR-SkN & 938.0 & \cellcolor[rgb]{ .573,  .816,  .314}\textbf{1023.1} & \cellcolor[rgb]{ .573,  .816,  .314}\textbf{949.5} & 1001.7 & 962.1 & \textbf{984.5} & 915.5 & \cellcolor[rgb]{ .573,  .816,  .314}\textbf{967.8} & \cellcolor[rgb]{ .573,  .816,  .314}\textbf{5.14} \\
    RE-RR-tN & 938.1 & 1025.3 & 954.2 & 1004.5 & 962.0 & 989.0 & 913.6 & 969.5 & 7.71 \\
    RE-RK-NN & 965.4 & \textit{1069.3} & 995.7 & 1013.0 & 996.9 & 1023.5 & 923.2 & 998.1 & 23.86 \\
    RE-RK-SkN & 946.3 & 1035.4 & 970.3 & 1003.6 & 967.7 & 999.3 & 912.2 & 976.4 & 13.43 \\
    RE-RK-tN & 946.5 & 1036.1 & 973.7 & \cellcolor[rgb]{ .573,  .816,  .314}\textbf{999.3} & 970.0 & 1013.1 & 912.1 & 978.7 & 13.14 \\
    RE-RV-RR-NN & 936.7 & 1062.4 & 973.7 & 1009.5 & 979.0 & 1047.7 & 934.6 & 992.0 & 18.71 \\
    RE-RV-RR-SkN & 940.6 & 1028.6 & 953.7 & 1002.0 & 961.5 & 987.4 & \cellcolor[rgb]{ .573,  .816,  .314}\textbf{906.0} & \textbf{968.5} & \textbf{5.57} \\
    RE-RV-RR-tN & 939.1 & 1034.1 & 955.6 & 1004.5 & 964.1 & 992.2 & 910.4 & 971.4 & 9.14 \\
    RE-RV-RK-NN & 960.4 & 1067.9 & 981.3 & 1015.4 & 995.4 & 1036.3 & 941.8 & 999.8 & 24.14 \\
    RE-RV-RK-SkN & 940.6 & 1027.6 & 954.7 & 1000.5 & 970.6 & \cellcolor[rgb]{ .573,  .816,  .314}\textbf{984.2} & 910.7 & 969.8 & 6.86 \\
    RE-RV-RK-tN & 945.3 & 1033.6 & 961.2 & 1004.2 & 970.7 & 990.4 & 911.1 & 973.8 & 11.71 \\
    RE-RR-RK-NN & 949.3 & 1060.7 & 967.3 & 1008.7 & 974.3 & 1039.4 & 927.7 & 989.6 & 19.14 \\
    RE-RR-RK-SkN & 934.7 & \textbf{1023.7} & 954.5 & \textbf{999.7} & 965.4 & 988.4 & 913.6 & 968.6 & \textbf{5.57} \\
    RE-RR-RK-tN & 942.2 & 1033.6 & \textbf{950.0} & 1003.2 & \textbf{961.1} & 992.4 & \textbf{909.6} & 970.3 & 7.29 \\
    RE-RV-RR-RK-NN & 938.7 & 1068.2 & 976.6 & 1007.5 & 983.3 & 1042.0 & 941.1 & 993.9 & 20.71 \\
    RE-RV-RR-RK-SkN & 938.3 & 1032.2 & 953.3 & 1001.9 & \cellcolor[rgb]{ .573,  .816,  .314}\textbf{957.3} & 988.5 & 910.9 & 968.9 & \textbf{5.57} \\
    RE-RV-RR-RK-tN & 934.2 & 1033.7 & 956.8 & 1001.0 & 961.7 & 996.9 & 910.8 & 970.7 & 7.43 \\
    \bottomrule
    \multicolumn{10}{p{55em}}{Note: For individual models, green highlight indicates the favoured model, bold indicates the 2nd ranked model, bold italics indicates the least favoured model, italics indicates the 2nd lowest ranked model, in each column.} \\     
    \end{tabular}%
}
\end{center}
  \label{tab:addlabel}%
\end{table}%

The other special case of the joint loss function of VaR and ES forecasts based on the asymmetric Laplace (AL) log score by \citet{Taylor2019} in equation (\ref{AL_log_score}) across models and market indices is presented in Table \ref{table:ALD_25pct} for 2.5\% forecast level and Table \ref{table:ALD_1pct} for 1\% forecast level, respectively. The proposed RE-SkN type models also produce the smallest average of the AL log scores (highlighted in green), i.e. for 5 out of 7 market indices, either for the 2.5\% or 1\% forecast levels. Some models in this class type also have the second lowest AL log score average.

The RE-RV-RR-RK-SkN model consistently generates the lowest joint loss and ranks first, on average, for both forecasting levels. For the 2.5\% forecasting level, the second best models with the smallest joint loss values are the RE-RR-SkN, RE-RR-RK-SkN, and RE-RV-RK-SkN models, whereas for 1\% forecasting level, the second best model is the RE-RR-SkN model. 

The comparison of VaR and ES joint loss functions using two different score functions, one by equation (\ref{VaRES_jointloss_specific}) and another one by equation (\ref{AL_log_score}), shows the superiority of the RE-SkN type models. It is also evident that the choice of the RRSS variable, either individually or jointly with the other two realized measures, can improve the tail-risk forecast performance of the RE-SkN type models. 

% Taylor (2019) AL loss is a special case of Fissler  and  Ziegel (2016).

\begin{table}[H]
  \begin{center}
  \captionsetup{font=normalsize, skip=0pt, justification= centering}
  \caption{Joint Loss for VaR and ES forecasts based on AL log score in equation (\ref{AL_log_score}); $\alpha=2.5\%$}
  \label{table:ALD_25pct}
  \resizebox{1.1\textwidth}{!}{
    \begin{tabular}{lccccccccc}
    \toprule
    \multicolumn{1}{c}{\textbf{Model}} & \textbf{ASX200} & \textbf{DAX} & \textbf{FTSE} & \textbf{HK} & \textbf{NASDAQ} & \textbf{SMI} & \textbf{S\&P500} & {\textbf{Mean loss}} & {\textbf{Mean Rank}} \\
    \midrule
    GARCH-t & \textit{1.901} & 2.209 & \textit{1.934} & 2.130 & 2.013 & 2.033 & 1.848 & \textit{2.010} & \textit{26.00} \\
    EGARCH-t & 1.837 & 2.207 & 1.898 & 2.099 & 1.961 & 2.057 & 1.775 & 1.976 & 24.00 \\
    GJR-t & 1.873 & 2.222 & 1.882 & 2.081 & 1.972 & 2.045 & 1.785 & 1.980 & 24.43 \\
    GARCH-t-HS & 1.823 & 2.172 & 1.892 & 2.119 & 1.974 & 2.009 & 1.817 & 1.972 & 21.57 \\
    GARCH-t-EVT & \textit{\textbf{2.317}} & \textit{\textbf{2.499}} & \textit{\textbf{2.280}} & \textit{\textbf{2.340}} & \textit{\textbf{2.093}} & \textit{\textbf{2.481}} & \textit{\textbf{2.030}} & \textit{\textbf{2.291}} & \textit{\textbf{30.00}} \\
    CARE-SAV & 1.831 & 2.170 & 1.892 & \textit{2.141} & 1.996 & 2.079 & \textit{1.870} & 1.997 & 24.00 \\
    RG-RV-tN & 1.762 & \textit{2.273} & 1.919 & 2.080 & 1.984 & 2.043 & 1.829 & 1.984 & 24.43 \\
    RG-RR-tN & 1.759 & 2.236 & 1.874 & 2.052 & 1.943 & 2.079 & 1.770 & 1.959 & 19.29 \\
    RG-RK-tN & 1.808 & 2.255 & 1.925 & 2.074 & \textit{2.035} & \textit{2.080} & 1.796 & 1.996 & \textit{26.00} \\
    RE-RV-NN & 1.762 & 2.216 & 1.870 & 2.067 & 1.931 & 2.020 & 1.769 & 1.948 & 19.14 \\
    RE-RV-SkN & 1.749 & 2.169 & 1.830 & 2.057 & 1.911 & \cellcolor[rgb]{ .573,  .816,  .314}\textbf{1.965} & 1.716 & 1.914 & 7.43 \\
    RE-RV-tN & 1.757 & 2.183 & 1.835 & 2.064 & 1.914 & 1.976 & 1.717 & 1.921 & 12.14 \\
    RE-RR-NN & 1.756 & 2.202 & 1.852 & \cellcolor[rgb]{ .573,  .816,  .314}\textbf{2.050} & 1.914 & 2.033 & 1.716 & 1.932 & 13.43 \\
    RE-RR-SkN & 1.753 & 2.168 & \textbf{1.816} & 2.057 & 1.903 & 1.977 & 1.711 & \textbf{1.912} & 5.86 \\
    RE-RR-tN & 1.749 & 2.170 & 1.826 & 2.091 & 1.902 & 1.976 & 1.716 & 1.918 & 9.14 \\
    RE-RK-NN & 1.791 & 2.210 & 1.876 & 2.067 & 1.945 & 2.009 & 1.725 & 1.946 & 19.00 \\
    RE-RK-SkN & 1.779 & 2.171 & 1.840 & 2.057 & 1.919 & 2.012 & 1.705 & 1.926 & 12.29 \\
    RE-RK-tN & 1.776 & 2.172 & 1.844 & 2.067 & 1.928 & 2.067 & 1.716 & 1.939 & 16.57 \\
    RE-RV-RR-NN & 1.753 & 2.210 & 1.855 & 2.058 & 1.914 & 2.030 & 1.741 & 1.937 & 15.43 \\
    RE-RV-RR-SkN & 1.761 & 2.181 & 1.823 & 2.060 & 1.904 & 1.977 & 1.705 & 1.916 & 9.00 \\
    RE-RV-RR-tN & 1.754 & 2.185 & 1.826 & 2.082 & 1.904 & 1.979 & \textbf{1.703} & 1.919 & 10.71 \\
    RE-RV-RK-NN & 1.775 & 2.215 & 1.867 & 2.065 & 1.933 & 2.010 & 1.750 & 1.945 & 18.71 \\
    RE-RV-RK-SkN & 1.751 & \textbf{2.166} & 1.822 & 2.057 & 1.911 & \textbf{1.966} & 1.711 & \textbf{1.912} & \textbf{5.43} \\
    RE-RV-RK-tN & 1.762 & 2.182 & 1.834 & 2.071 & 1.915 & 1.969 & 1.709 & 1.920 & 12.00 \\
    RE-RR-RK-NN & 1.762 & 2.213 & 1.842 & 2.058 & 1.910 & 2.018 & 1.731 & 1.933 & 15.00 \\
    RE-RR-RK-SkN & \cellcolor[rgb]{ .573,  .816,  .314}\textbf{1.741} & \cellcolor[rgb]{ .573,  .816,  .314}\textbf{2.165} & 1.823 & \textbf{2.052} & 1.909 & 1.981 & 1.714 & \textbf{1.912} & \textbf{5.43} \\
    RE-RR-RK-tN & 1.755 & 2.182 & \cellcolor[rgb]{ .573,  .816,  .314}\textbf{1.815} & 2.087 & \textbf{1.901} & 1.982 & \cellcolor[rgb]{ .573,  .816,  .314}\textbf{1.698} & 1.917 & 9.14 \\
    RE-RV-RR-RK-NN & 1.748 & 2.219 & 1.857 & 2.056 & 1.926 & 2.021 & 1.750 & 1.940 & 15.14 \\
    RE-RV-RR-RK-SkN & 1.748 & 2.180 & 1.821 & 2.059 & \cellcolor[rgb]{ .573,  .816,  .314}\textbf{1.894} & 1.971 & 1.704 & \cellcolor[rgb]{ .573,  .816,  .314}\textbf{1.911} & \cellcolor[rgb]{ .573,  .816,  .314}\textbf{5.14} \\
    RE-RV-RR-RK-tN & \textbf{1.743} & 2.179 & 1.828 & 2.081 & 1.901 & 1.979 & 1.710 & 1.918 & 8.86 \\
    \bottomrule
    \multicolumn{10}{p{55em}}{Note: For individual models, green highlight indicates the favoured model, bold indicates the 2nd ranked model, bold italics indicates the least favoured model, italics indicates the 2nd lowest ranked model, in each column.} \\     
    \end{tabular}%
}
\end{center}
  \label{tab:addlabel}%
\end{table}%

\begin{table}[H]
  \begin{center}
  \captionsetup{font=normalsize, skip=0pt, justification= centering}
  \caption{Joint Loss for VaR and ES forecasts based on  AL log score in equation (\ref{AL_log_score}); $\alpha=1\%$}
  \label{table:ALD_1pct}
  \resizebox{1.1\textwidth}{!}{
    \begin{tabular}{lccccccccc}
    \toprule
    \multicolumn{1}{c}{\textbf{Model}} & \textbf{ASX200} & \textbf{DAX} & \textbf{FTSE} & \textbf{HK} & \textbf{NASDAQ} & \textbf{SMI} & \textbf{S\&P500} & {\textbf{Mean loss}} & {\textbf{Mean Rank}} \\
    \midrule
    GARCH-t & \textit{2.064} & 2.365 & 2.072 & 2.293 & 2.110 & 2.246 & 1.980 & 2.161 & \textit{23.86} \\
    EGARCH-t & 2.027 & 2.441 & 2.094 & 2.261 & 2.051 & 2.347 & 1.903 & 2.161 & \textit{23.86} \\
    GJR-t & 2.029 & 2.439 & 2.025 & 2.231 & 2.063 & 2.335 & 1.900 & 2.146 & 21.71 \\
    GARCH-t-HS & 1.960 & 2.347 & 2.056 & 2.266 & 2.049 & 2.230 & 1.968 & 2.125 & 20.29 \\
    GARCH-t-EVT & \textit{\textbf{2.364}} & \textit{\textbf{2.797}} & \textit{\textbf{2.591}} & \textit{\textbf{2.434}} & \textit{\textbf{2.358}} & \textit{\textbf{3.046}} & \textit{\textbf{2.190}} & \textit{\textbf{2.540}} & \textit{\textbf{30.00}} \\
    CARE-SAV & 1.962 & 2.321 & 2.070 & \textit{2.340} & 2.128 & 2.296 & \textit{2.025} & 2.163 & 23.00 \\
    RG-RV-tN & \cellcolor[rgb]{ .573,  .816,  .314}\textbf{1.885} & \textit{2.472} & \textit{2.102} & 2.262 & 2.139 & 2.354 & 1.998 & 2.173 & 23.57 \\
    RG-RR-tN & \textbf{1.888} & 2.404 & 2.021 & 2.195 & 2.060 & \textit{2.405} & 1.894 & 2.124 & 15.57 \\
    RG-RK-tN & 1.956 & 2.437 & 2.094 & 2.232 & \textit{2.225} & 2.334 & 1.950 & \textit{2.175} & 23.29 \\
    RE-RV-NN & 1.911 & 2.441 & 2.046 & 2.267 & 2.106 & 2.372 & 1.960 & 2.158 & \textit{23.86} \\
    RE-RV-SkN & 1.900 & 2.303 & 1.965 & 2.202 & 2.032 & \textbf{2.189} & 1.828 & 2.060 & 8.57 \\
    RE-RV-tN & 1.903 & 2.327 & 1.972 & 2.205 & 2.038 & 2.211 & 1.816 & 2.067 & 11.29 \\
    RE-RR-NN & 1.902 & 2.401 & 2.021 & 2.212 & 2.057 & 2.381 & 1.821 & 2.114 & 16.71 \\
    RE-RR-SkN & 1.898 & \cellcolor[rgb]{ .573,  .816,  .314}\textbf{2.294} & \textbf{1.948} & 2.197 & 2.016 & 2.189 & 1.827 & \textbf{2.053} & 6.00 \\
    RE-RR-tN & 1.895 & \textbf{2.295} & 1.960 & 2.233 & 2.013 & 2.192 & 1.824 & 2.059 & 8.14 \\
    RE-RK-NN & 1.973 & 2.439 & 2.069 & 2.249 & 2.105 & 2.311 & 1.849 & 2.142 & 21.71 \\
    RE-RK-SkN & 1.924 & 2.325 & 1.996 & 2.215 & 2.023 & 2.250 & 1.816 & 2.078 & 12.71 \\
    RE-RK-tN & 1.926 & 2.328 & 2.003 & 2.201 & 2.036 & 2.323 & 1.821 & 2.091 & 13.71 \\
    RE-RV-RR-NN & 1.894 & 2.430 & 2.022 & 2.230 & 2.054 & 2.372 & 1.880 & 2.126 & 17.71 \\
    RE-RV-RR-SkN & 1.908 & 2.312 & 1.958 & 2.199 & 2.008 & 2.199 & \cellcolor[rgb]{ .573,  .816,  .314}\textbf{1.795} & 2.054 & 6.00 \\
    RE-RV-RR-tN & 1.898 & 2.336 & 1.963 & 2.225 & 2.015 & 2.203 & \textbf{1.808} & 2.064 & 9.14 \\
    RE-RV-RK-NN & 1.976 & 2.443 & 2.041 & 2.249 & 2.101 & 2.340 & 1.905 & 2.151 & 23.29 \\
    RE-RV-RK-SkN & 1.909 & 2.307 & 1.958 & \textbf{2.195} & 2.031 & \cellcolor[rgb]{ .573,  .816,  .314}\textbf{2.185} & 1.809 & 2.056 & 6.00 \\
    RE-RV-RK-tN & 1.927 & 2.320 & 1.975 & 2.207 & 2.036 & 2.199 & 1.811 & 2.068 & 11.14 \\
    RE-RR-RK-NN & 1.939 & 2.427 & 2.002 & 2.227 & 2.051 & 2.345 & 1.860 & 2.122 & 18.00 \\
    RE-RR-RK-SkN & 1.891 & 2.297 & 1.962 & \cellcolor[rgb]{ .573,  .816,  .314}\textbf{2.187} & 2.021 & 2.202 & 1.820 & 2.054 & \textbf{5.86} \\
    RE-RR-RK-tN & 1.918 & 2.320 & \cellcolor[rgb]{ .573,  .816,  .314}\textbf{1.944} & 2.222 & 2.012 & 2.203 & 1.809 & 2.061 & 8.29 \\
    RE-RV-RR-RK-NN & 1.896 & 2.450 & 2.027 & 2.228 & 2.076 & 2.354 & 1.893 & 2.132 & 19.57 \\
    RE-RV-RR-RK-SkN & 1.898 & 2.312 & 1.957 & 2.196 & \cellcolor[rgb]{ .573,  .816,  .314}\textbf{1.991} & 2.195 & 1.808 & \cellcolor[rgb]{ .573,  .816,  .314}\textbf{2.051} & \cellcolor[rgb]{ .573,  .816,  .314}\textbf{4.43} \\
    RE-RV-RR-RK-tN & 1.891 & 2.316 & 1.966 & 2.205 & \textbf{2.007} & 2.219 & 1.813 & 2.059 & 7.43 \\
    \bottomrule
    \multicolumn{10}{p{55em}}{Note: For individual models, green highlight indicates the favoured model, bold indicates the 2nd ranked model, bold italics indicates the least favoured model, italics indicates the 2nd lowest ranked model, in each column.} \\     
    \end{tabular}%
}
\end{center}
  \label{tab:addlabel}%
\end{table}%

\subsection{Model Confidence Set}

The joint loss function based on the AL log score function (equation (\ref{AL_log_score})) for both 2.5\% and 1\% forecasting levels are further incorporated as the loss function in the calculation of the MCS \citep{HansenEtAl2011}. The tests are conducted by using both R and SQ methods at 90\% and 75\% confidence levels. The result of the tests are presented in Table \ref{table:MCS}. The models that consistently enter the MCS for both MCS methods and across market indices are highligted in green. Overall, the proposed extensions of the realized EGARCH models (the RE-SkN and RE-tN type models), demonstrate a superior performance than the competing models. The RE-SkN and RE-tN type models are more likely to be included in the R and SQ MCS methods, both at the 90\% and 75\% confidence levels. They are also more favored than the original realized EGARCH models that are estimated using the proposed adaptive MCMC in this paper (the RE-NN type models).

\begin{table}[H]
  \begin{center}
  \captionsetup{font=normalsize, skip=0pt, justification= centering}
  \caption{75\% and 90\% Model confidence set (MCS) with R and SQ methods}
  \label{table:MCS}
  \resizebox{0.75\textwidth}{!}{
    \begin{tabular}{lcccccccc}
    \toprule
    \multicolumn{1}{c}{\multirow{2}[6]{*}{\textbf{Model}}} & \multicolumn{4}{c}{\textbf{$\alpha$=2.5\%}} & \multicolumn{4}{c}{\textbf{$\alpha$=1\%}} \\
\cmidrule{2-9}          & \multicolumn{2}{c}{\textbf{90\% MCS}} & \multicolumn{2}{c}{\textbf{75\% MCS}} & \multicolumn{2}{c}{\textbf{90\% MCS}} & \multicolumn{2}{c}{\textbf{75\% MCS}} \\
\cmidrule{2-9}          & \multicolumn{1}{c}{\textbf{R}} & \multicolumn{1}{c}{\textbf{SQ}} & \multicolumn{1}{c}{\textbf{R}} & \multicolumn{1}{c}{\textbf{SQ}} & \multicolumn{1}{c}{\textbf{R}} & \multicolumn{1}{c}{\textbf{SQ}} & \multicolumn{1}{c}{\textbf{R}} & \multicolumn{1}{c}{\textbf{SQ}} \\
    \midrule
    GARCH-t & 4     & 2     & 1     & 2     & 6     & 5     & 4     & 3 \\
    EGARCH-t & 6     & 5     & 5     & 3     & 6     & 5     & 5     & 4 \\
    GJR-t & 5     & 3     & 4     & 2     & 6     & 5     & 6     & 5 \\
    GARCH-t-HS & 4     & 4     & 3     & 3     & 6     & 5     & 5     & 4 \\
    GARCH-t-EVT & 2     & 2     & 1     & 1     & 4     & 2     & 3     & 2 \\
    CARE-SAV & 5     & 4     & 4     & 2     & 4     & 4     & 4     & 2 \\
    RG-RV-tN & 7     & 3     & 4     & 3     & 6     & 4     & 4     & 2 \\
    RG-RR-tN & 6     & 5     & 6     & 3     & 6     & 5     & 6     & 4 \\
    RG-RK-tN & 6     & 2     & 3     & 2     & 6     & 6     & 5     & 3 \\
    RE-RV-NN & 7     & 4     & 6     & 3     & 6     & 5     & 4     & 2 \\
    RE-RV-SkN & 7     & 7     & 7     & 7     & 7     & 6     & 7     & 6 \\
    RE-RV-tN & 7     & 5     & 7     & 5     & 7     & 6     & 7     & 6 \\
    RE-RR-NN & 7     & 6     & 7     & 5     & 7     & 6     & 7     & 5 \\
    RE-RR-SkN & 7     & 7     & 7     & 7     & 6     & 6     & 6     & 6 \\
    RE-RR-tN & 7     & 6     & 6     & 6     & 6     & 6     & 5     & 6 \\
    RE-RK-NN & 7     & 5     & 6     & 3     & 6     & 5     & 4     & 2 \\
    RE-RK-SkN & 7     & 7     & 7     & 7     & 7     & 6     & 7     & 6 \\
    RE-RK-tN & 6     & 5     & 6     & 4     & 7     & 5     & 7     & 4 \\
    RE-RV-RR-NN & 7     & 5     & 6     & 4     & 6     & 5     & 5     & 4 \\
    RE-RV-RR-SkN & \cellcolor[rgb]{ .573,  .816,  .314}\textbf{7} & \cellcolor[rgb]{ .573,  .816,  .314}\textbf{7} & \cellcolor[rgb]{ .573,  .816,  .314}\textbf{7} & \cellcolor[rgb]{ .573,  .816,  .314}\textbf{7} & \cellcolor[rgb]{ .573,  .816,  .314}\textbf{7} & \cellcolor[rgb]{ .573,  .816,  .314}\textbf{7} & \cellcolor[rgb]{ .573,  .816,  .314}\textbf{7} & \cellcolor[rgb]{ .573,  .816,  .314}\textbf{7} \\
    RE-RV-RR-tN & 7     & 6     & 7     & 6     & 7     & 6     & 6     & 6 \\
    RE-RV-RK-NN & 7     & 4     & 6     & 3     & 6     & 5     & 4     & 3 \\
    RE-RV-RK-SkN & \cellcolor[rgb]{ .573,  .816,  .314}\textbf{7} & \cellcolor[rgb]{ .573,  .816,  .314}\textbf{7} & \cellcolor[rgb]{ .573,  .816,  .314}\textbf{7} & \cellcolor[rgb]{ .573,  .816,  .314}\textbf{7} & \cellcolor[rgb]{ .573,  .816,  .314}\textbf{7} & \cellcolor[rgb]{ .573,  .816,  .314}\textbf{7} & \cellcolor[rgb]{ .573,  .816,  .314}\textbf{7} & \cellcolor[rgb]{ .573,  .816,  .314}\textbf{7} \\
    RE-RV-RK-tN & \cellcolor[rgb]{ .573,  .816,  .314}\textbf{7} & \cellcolor[rgb]{ .573,  .816,  .314}\textbf{7} & \cellcolor[rgb]{ .573,  .816,  .314}\textbf{7} & \cellcolor[rgb]{ .573,  .816,  .314}\textbf{7} & \cellcolor[rgb]{ .573,  .816,  .314}\textbf{7} & \cellcolor[rgb]{ .573,  .816,  .314}\textbf{7} & \cellcolor[rgb]{ .573,  .816,  .314}\textbf{7} & \cellcolor[rgb]{ .573,  .816,  .314}\textbf{7} \\
    RE-RR-RK-NN & 7     & 5     & 6     & 5     & 6     & 5     & 5     & 5 \\
    RE-RR-RK-SkN & 7     & 6     & 7     & 6     & 6     & 6     & 6     & 6 \\
    RE-RR-RK-tN & 7     & 6     & 6     & 6     & 7     & 6     & 7     & 6 \\
    RE-RV-RR-RK-NN & 7     & 5     & 6     & 4     & 6     & 5     & 5     & 4 \\
    RE-RV-RR-RK-SkN & \cellcolor[rgb]{ .573,  .816,  .314}\textbf{7} & \cellcolor[rgb]{ .573,  .816,  .314}\textbf{7} & \cellcolor[rgb]{ .573,  .816,  .314}\textbf{7} & \cellcolor[rgb]{ .573,  .816,  .314}\textbf{7} & \cellcolor[rgb]{ .573,  .816,  .314}\textbf{7} & \cellcolor[rgb]{ .573,  .816,  .314}\textbf{7} & \cellcolor[rgb]{ .573,  .816,  .314}\textbf{7} & \cellcolor[rgb]{ .573,  .816,  .314}\textbf{7} \\
    RE-RV-RR-RK-tN & \cellcolor[rgb]{ .573,  .816,  .314}\textbf{7} & \cellcolor[rgb]{ .573,  .816,  .314}\textbf{7} & \cellcolor[rgb]{ .573,  .816,  .314}\textbf{7} & \cellcolor[rgb]{ .573,  .816,  .314}\textbf{7} & \cellcolor[rgb]{ .573,  .816,  .314}\textbf{7} & \cellcolor[rgb]{ .573,  .816,  .314}\textbf{7} & \cellcolor[rgb]{ .573,  .816,  .314}\textbf{7} & \cellcolor[rgb]{ .573,  .816,  .314}\textbf{7} \\

    \bottomrule
    \multicolumn{9}{p{34em}}{Note: For each MCS method, green indicates favored models for all market indices.} \\  
    \end{tabular}%
}
\end{center}
  \label{tab:addlabel}%
\end{table}% 

The result in Table \ref{table:MCS} also indicates that there is a tendency that the realized EGARCH models employing multiple realized measures have a more favorable performance than those using single realized measure. In particular, there are 3 RE-SkN type models and 2 RE-tN type models using multiple realized measures that are consistently included in the MCS using both the R and SQ methods at both confidence levels, i.e. the RE-RV-RR-SkN, RE-RV-RK-SkN, RE-RV-RR-RK-SkN, RE-RV-RK-tN, and RE-RV-RR-RK-tN models. The RE-RV-RR-RK-SkN is also the most preferred model based on the VaR and ES joint loss function values using the AL log score function (equation (\ref{AL_log_score})). Meanwhile, the RE-RR-SkN model, which is the second best model based on the joint loss function values using the loss functions in equation (\ref{VaRES_jointloss_specific}) and equation (\ref{AL_log_score}), is all included in the MCS for 7 market indices when the MCS incorporates the 2.5\% VaR and ES levels. Although the RE-RR-SkN model is included in the MCS for only 6 market indices when using the 1\% tail risk forecasting, the performance of this model overall is still highly favorable.

\section{Conclusion}
\label{sec:conclusion}

In this paper, we propose a Bayesian approach to estimate the realized EGARCH model for the purpose of VaR and ES forecasting. The original realized EGARCH model is extended by employing Student-\textit{t} and skewed Student-\textit{t} distributions for the observation equation errors (RE-tN and RE-SkN type models). The result of the simulation study demonstrates the advantage of the MCMC estimators over maximum likelihood estimators in terms of unbiasedness and efficiency.
 
In the empirical study using 7 market indices, the sub-sampled realized variance, the sub-sampled realized range, as well as the realized kernel are chosen to improve the out-of-sample forecasting performance of the models. The empirical study result shows that the choice of the return equation error distribution is found to be consequential. The proposed RE-SkN class models have a superior tail risk forecasting performance in most cases than the standard GARCH-type models, CARE models, realized GARCH models employing the Student-\textit{t} distribution for the observation equation errors. The proposed RE-SkN class models consistently produce conservative and accurate violation rates and sufficient risk coverage, are generally less likely to be rejected in both VaR and ES backtests, produce the lowest joint loss function values, and most of the time are included in the model confidence set. 

This paper also finds that both RE-SkN and and RE-tN type models are more favorable than than the original realized EGARCH model using the Gaussian distribution estimated under our proposed Bayesian framework (RE-NN type models). With respect to the choice of realized measures, the sub-sampled realized range (either employed individually or jointly with sub-sampled realized variance and/or realized kernel) is found to be crucial in improving the accuracy of the tail risk forecasts. 

We conclude that the RE-SkN and RE-tN type models incorporating sub-sampled realized range should be considered in tail risk forecasting. This will help financial institutions to improve capital allocation efficiency and allow them to to have larger profit maximization opportunities while at the same time be able to maintain safe goals in accordance to the Basel Capital Accords. 

This paper can be extended by: allowing other distributions for both the return and measurement equation errors, e.g. by using asymmetric Laplace or two-sided Weibull distributions; incorporating a higher number of different realized measures; using other data frequencies and sub-sampling frequency for the realized measures; and employing other MCMC algorithms.  Another area to explore is to develop a variable selection and shrinkage method for selecting realized measures in the realized EGARCH models under a Bayesian framework.  

\clearpage

% Make appendix
% \newpage
% \appendix
% \section{Appendix: the convergence of MCMC}
\end{onehalfspace}

\newpage
% \bibliographystyle{apacite}
% \bibliographystyle{chicago}
% \bibliography{References}

\end{document}